\documentclass[twocolumn,aps,superscriptaddress,multicol,amsmath,amssymb]{revtex4-2}
\usepackage{amssymb}
\usepackage{amsmath}
\usepackage{graphicx}
\usepackage{epstopdf}
\usepackage{bm}
\usepackage{gensymb}
\usepackage{soul}
\usepackage{sidecap}
\usepackage[normalem]{ulem}
\usepackage{times}

\usepackage{float}
\usepackage[caption = false]{subfig}
\usepackage{subfig}
\usepackage{enumerate}
\usepackage{multirow}
\usepackage{tabularx}
\usepackage{array}
\usepackage{url}
\usepackage{slantsc}
\usepackage{lmodern}
\newcommand\redout{\bgroup\markoverwith{\textcolor{red}{\rule[.5ex]{2pt}{0.4pt}}}\ULon}

\usepackage[colorlinks=true,citecolor=blue]{hyperref}
\hypersetup{colorlinks=true,citecolor=blue,linkcolor=red,urlcolor=blue}

\usepackage{hyperref,cleveref}
\newcommand{\be}{\begin{equation}}
	\newcommand{\ee}{\end{equation}}
\newcommand{\bk}{{{\bf{k}}}}
\newcommand{\bq}{{{\bf{q}}}}

\newcommand{\bA}{{{\bf{A}}}}
\newcommand{\bJ}{{{\bf{J}}}}

\newcommand{\br}{{{\bf{r}}}}

\newcommand{\bea}{\begin{eqnarray}}
	\newcommand{\eea}{\end{eqnarray}}

\newcommand{\bd}{\begin{displaymath}}
	\newcommand{\ed}{\end{displaymath}}
\newcommand{\ba}{\begin{array}}
	\newcommand{\ea}{\end{array}}
\newcommand{\bi}{\begin{itemize}}
	\newcommand{\ei}{\end{itemize}}
\newcommand{\bc}{\begin{center}}
	\newcommand{\ec}{\end{center}}
\newcommand{\bfl}{\begin{flushleft}}
	\newcommand{\efl}{\end{flushleft}}
\newcommand{\bfr}{\begin{flushright}}
	\newcommand{\efr}{\end{flushright}}

\newcommand{\mi}{\rm i}
\def\br{{\bf r}}
\def\bk{{\bf k}} \def\bq{{\bf q}}  
  \def\bd{{\bf d}}

\def\6{\partial}

\def\={\!\!\!&=&\!\!\!}
\def\+{\!\!\!&&\!\!\!+~}
\def\-{\!\!\!&&\!\!\!-~}

\graphicspath{{.}{./Figs/}}
\usepackage{color}
\usepackage[dvipsnames]{xcolor}
\usepackage{xcolor}
\begin{document}
	\title{Microscopic Insights into London Penetration Depth: Application to  CeCoIn$^{}_{5}$} 
	\author{Mehdi Biderang}
	\affiliation{Department of Physics, University of Toronto, 60 St. George Street, Toronto, Ontario, M5S 1A7, Canada}
	\author{Jeehoon Kim}
	\affiliation{Department of Physics, POSTECH, Pohang, Gyeongbuk 790-784, Korea}
    \author{Reza Molavi}
	\affiliation{Department of Electrical and Computer Engineering, University of British Columbia, Vancouver, British Columbia, Canada}
     \affiliation{D-Wave Systems Inc., Burnaby, British Columbia, Canada}
	\author{Alireza Akbari}
	\affiliation{Max Planck Institute for the Chemical Physics of Solids, 01187 Dresden, Germany}
	\affiliation{Asia Pacific Center for Theoretical Physics (APCTP), Pohang, Gyeongbuk, 790-784, Korea}
    \affiliation{Max Planck POSTECH Center for Complex Phase Materials, POSTECH, Pohang, Gyeongbuk 790-784, Korea}
	%
	\date{\today}
	%
	\begin{abstract}
		
		We propose a comprehensive theoretical  formulation of magnetic penetration depth, $\lambda(T)$, based on the microscopic calculations for a general superconducting gap symmetry. 
		Our findings admit the significant role of band structure and Fermi surface topology together with the symmetry of superconducting order parameter.
		We employ  our findings pertaining to the heavy-fermion superconductor CeCoIn$_5$ to explore both local and non-local behaviors in response to an external magnetic field across varying temperatures.
        Our calculations in the low-temperature regime offer compelling macroscopic evidence of the nodal character within the superconducting state with $d_{x^2-y^2}$ symmetry. Furthermore, our findings align with the characteristics of London-type superconductivity, holding significant implications for upcoming experiments.
	\end{abstract}
	%
	\maketitle
	%
	\section{Introduction}
	The discovery of the  unconventional superconductivity has opened up exciting avenues of research, offering unprecedented insights into the fundamental properties of matter~\cite{Muller_Bednorz_1986,Monthoux_2007,Norman_2011}. 
	The superconducting properties  have been extensively investigated using various theoretical and experimental techniques, shedding light on its electronic structure, pairing mechanism, and gap symmetry~\cite{Scalapino_RevModPhys_1012, Mineev-Samokhin-1999}.
	Of particular significance is the measurement of the London penetration depth, a fundamental property that delineates a superconductor's response to an applied magnetic field~\cite{Prozorov_2006,Abrikosov_1975,Eliashberg_1991,Tinkham_2004,Wen_PRB_2009,Prozonov_PRB_2021}. 
	The experimental studies employing diverse techniques, such as muon spin rotation ($\mu$SR), specific heat measurements, scanning tunneling microscopy (STM), and magnetic force microscopy (MFM) have explored the  penetration depth~\cite{Barford_1988,Rossel_PRB_1990,Roseman_2001,Keller_PRB_2006,JeeHoonKim_2012,Eltschka_2015,Coffey_JAP_2002}. 
    To illustrate the significance of London penetration depth in practical application, it is worthwhile to recognize that superconducting quantum bits (qubits) are typically designed to possess specific body inductance comprised of two major components, geometric and kinetic inductance. The latter term is heavily impacted by the London penetration depth further underscoring the necessity of its accurate modeling to predict individual qubit performance in a large-scale quantum processor~\cite{Harris_PRB_2012}.
	The temperature-dependent behavior of the penetration depth and related superfluid density provides direct windows into the nature of the superconducting gap~\cite{Leggett_PRL_1997,Sigrist_PRB_2006,Chen_2013}.
	These investigations have revealed intriguing behavior, including a rapid suppression of the superfluid density with increasing temperature, unconventional power-law temperature dependencies, and the presence of nodes or anisotropic gap structures in the superconducting state~\cite{Guguchia_2019,Chun_PRL_2020,Welp_PRL_2021,Dzhumanov_2022}. 
	Consequently, it becomes a reflection of the interplay between the superconducting gap structure and the topology of the Fermi surface~\cite{Fletcher_2007,Prozonov_PRL_2015}.
	A comprehensive understanding of the penetration depth in both `specular' and `diffuse' regimes is crucial for investigating the 
	designing devices that utilize superconductors, such as high-speed electronics, magnetic resonance imaging (MRI) machines, or particle accelerators.
	\\

	The heavy-fermion compounds have risen as captivating systems that manifest unconventional superconductivity, pushing the boundaries of our comprehension beyond the confines of the conventional BCS (Bardeen-Cooper-Schrieffer) theory~\cite{Steglich-1979, Hewson-1993,  Pfleiderer_REvModPhys_2009}. Among these compounds, CeCoIn$_{5}$ has attracted considerable attention due to its intriguing superconducting properties and the presence of a complex superconducting gap structure~\cite{Petrovic_2001, Bianchi_PRL_2003,Paglione_PRL_2003,Sarrao_2007,Fisk_2012,AliYazdani_NatComm_2018}.
	It is a unique heavy-fermion compound that crystallizes in the tetragonal HoCoGa$_{5}$-type structure and exhibits superconductivity below a critical temperature of $T_{c}\approx 2.3$K~\cite{Petrovic_2001}. This compound is of particular interest because it resides in close proximity to a quantum critical point, where tiny perturbations can induce significant changes in its electronic properties. 
	This proximity to a quantum critical point hints at the possibility of unconventional Cooper pairing, emphasizing the intricate interplay between magnetism and superconductivity~\cite{Sidorov_PRL_2002, Izawa_PRL_2001}.
	These findings challenge conventional models and highlight the need for a comprehensive study to elucidate the superconducting gap function.
	Several hypotheses regarding its superconducting gap function have been explored, drawing from thermodynamic and transport properties. The evidence suggests a spin singlet gap with d-wave pairing. Initially, conflicting results hinted at both $d_{x^2-y^2}$ and     $d_{xy}$ gap symmetries~\cite{Greene_PRL_2008,Bianchi_PRL_2003}. However, a distinct spin resonance observed via inelastic neutron scattering strongly supports the $d_{x^2-y^2}$ symmetry~\cite{Stock_PRL_2008, PhysRevLett.101.187001,PhysRevB.86.134516}.
	Additionally, low-temperature field-angle-resolved specific heat measurements provided further confirmation of this symmetry~\cite{PhysRevLett.104.037002}, and subsequent Quasiparticle interference (QPI) measurements reinforced it~\cite{Akbari_PRB_2011,Allan_2013,AliYazdani_Nature_2013}.
	\\

    	In addition to providing strong evidence for the $d$-wave properties of CeCoIn$_5$, the most recent magnetic force microscopy  results have shown a peculiar power-law pattern in the penetration depth~\cite{JeeHooon_Kim_2020}.
        Inspired by these discoveries, our work presents a comprehensive theoretical analysis of the penetration depth in such materials, utilizing microscopic approaches.
        We take into account both local and non-local responses as we investigate the behaviour of penetration depth  across a broad temperature range. Our primary goal is to give a thorough explanation of its dependence on the superconducting gap symmetry. 
	In this regard,
	Section II  outlines the physical model used to describe the electronic band structure.
	In Section III, we introduce the response of a superconductor to an external electromagnetic field and establish a microscopic model to investigate the magnetic penetration depth within a superconducting context.
	Lastly, in Section IV, we consolidate our findings, presenting potential scenarios for the superconducting gap function in CeCoIn$_{5}$.
	\\

	\section{Model Hamiltonian of a heavy Fermion system}
\begin{figure}[t]
		\includegraphics[width=0.7\linewidth]{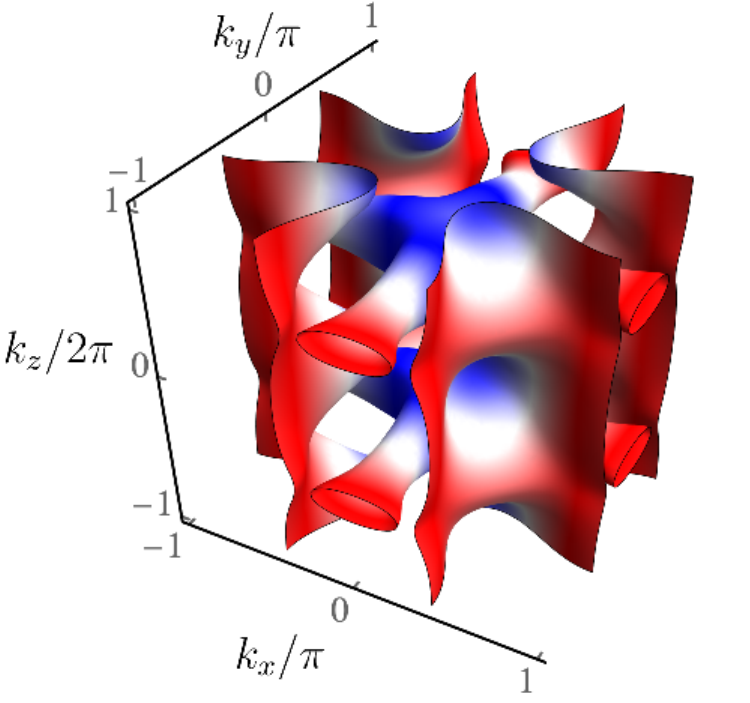}   %
		\caption{Calculated three-dimensional Fermi surfaces of  
			CeCoIn$^{}_{5}$ based on Anderson lattice model.  
		}
		\label{Fig:Fermi_Surface}
	\end{figure}

	The Anderson lattice model Hamiltonian, which takes into account the hybridization between the doubly spin-degenerate conduction and localized orbitals  can be represented as~\cite{Tanaka_JKPS_2006}: 
	%
	%
	\begin{equation}
		{\cal H}={\cal H}^{}_{0}+{\cal H}^{}_{\rm int},
	\end{equation}
	with the non-interacting part of Hamiltonian represented by
	%
	%
	\begin{equation}
		{\cal H}^{}_{0}
		= \sum^{}_{\bk,\sigma}
		\varepsilon^{c}_{\bk} c^{\dagger}_{\bk,\sigma}c^{}_{\bk,\sigma}
		+\varepsilon^{f}_{\bk} f^{\dagger}_{\bk,\sigma}f^{}_{\bk,\sigma}
		+\frac{1}{2}(V^{}_{\bk} c^{\dagger}_{\bk,\sigma}f^{}_{\bk,\sigma} + {\rm h.c.} ),
	\end{equation}
	and the on-site Coulomb interaction among the $f$ electrons given by
	%
	%
	\begin{equation}
		{\cal H}^{}_{\rm int}
		=\sum^{}_{\bk\bk'}
		U^{}_{f\!f}
		f^{\dagger}_{\bk,\uparrow}
		f^{}_{\bk,\downarrow}
		f^{\dagger}_{\bk',\uparrow}
		f^{}_{\bk',\downarrow}.
	\end{equation}
	%
	%
	Here the creation of a conduction (localized) electron of momentum $\bk$ and spin $\sigma$ is denoted by $c^{\dagger}_{\bk,\sigma}$ ($f^{\dagger}_{\bk,\sigma}$). 
	The hybridization between the lowest $4f$ doublet and conduction bands occurs through $V_{\mathbf{k}}$, encompassing the influence of spin-orbit coupling and crystal field effects.
	%
	%
	As the on-site Coulomb interaction $U_{f\!f}$ tends towards an extremely large value, the existence of doubly occupied $f$-states becomes energetically forbidden.
	Consequently, by introducing auxiliary bosons, the mean-field (MF) Hamiltonian of the quasiparticle bands can be expressed as follows:
	%
	%
	\begin{equation}
		{\cal H}^{}_{\rm MF}
		=\sum^{}_{\bk,s,\sigma}
		\xi^{}_{\bk,s}
		a^{\dagger}_{\bk,s,\sigma}
		a^{}_{\bk,s,\sigma},
		\label{Eq:MF_Ham}
	\end{equation}
	%
	%
	where $a^{\dagger}_{\bk,s,\sigma}$ is an operator of creation of a quasiparticle at band $s=\pm$ and spin $\sigma$.
	The dispersion of quasiarticles is given by
	%
	%
	\begin{equation}
		\xi^{}_{\bk,\pm}=
		\frac{1}{2}
		\Big[
		\varepsilon^{c}_{\bk}
		+
		\varepsilon^{f}_{\bk}
		\pm
		\sqrt{
			(\varepsilon^{c}_{\bk}
			-
			\varepsilon^{f}_{\bk})^2
			+
			V^{2}_{\bk}
		}	
		\Big].
		\label{Eq:Dispersion_quasiparticles}
	\end{equation}
	%
	%
	The effective hybridization between conduction and localized electrons is denoted by $\tilde{V}^{}_{\bk}= V^{}_{\bk}\sqrt{1-n_{f}}$, where $n^{}_f$ represents the occupancy of the localized electron states and ensures the exclusion of double occupancy.
	Extensive studies have demonstrated that the formation of the Fermi surface is solely attributed to the $s=-$ band, while the $s=+$ band resides above the Fermi energy.
    This phenomenon becomes evident in a three-dimensional arrangement within a tetragonal crystal lattice, with a distinct emphasis on the $C^{}_{4v}$ rotational symmetry within the $xy$-plane.
	By employing the  model detailed in Refs.~\cite{Tanaka_JKPS_2006,Akbari_PRB_2011},  we illustrates the Fermi surface of CeCoIn$_5$ in Fig.~\ref{Fig:Fermi_Surface}.
	%
	%
	Moreover, the superconducting state of the system can be described by
	\begin{equation}
		{\cal H}_{\rm SC}=
		\sum_{\bk,s}
		(
		\Delta^{}_{\bk}
		a^{\dagger}_{\bk,s,\uparrow}
		a^{\dagger}_{-\bk,s,\downarrow}
		+
		{\rm h.c.}
		),
		\label{Eq:Ham_Sc}
	\end{equation}
	%
	wherein $\Delta^{}_{\bk}$ represents the superconducting gap function in the singlet channel.
	Extensive research has established that the gap function symmetry in CeCoIn$^{}_5$ exhibits the characteristic nodal structure of $d$-wave pairing.
	%
	%
	
	%
	\section{Response of a superconductor in a weak electromagnetic filed}
	We investigate the response of a superconductor to an external weak electromagnetic field, which is significantly smaller than the critical magnetic fields.
	We focus our analysis on a semi-infinite three-dimensional  CeCoIn$_5$ superconductor  with a planar surface located at $z=0$, occupying the half-space ($z<0$).
	This system is exposed to a constant weak magnetic field (Meissner state) in $y$-direction.
	The current density $\mathbf{J}$ is connected to the vector potential $\mathbf{A}$ through the relation~\cite{Abrikosov_1975}
	%
	\begin{align}
		\begin{aligned}
			\mathbf{J}(\br,t,T)=-\int \mathbb{K}(\br-\br',t-t',T)\mathbf{A}(\mathbf{r}',t') d\br' dt',
		\end{aligned}
	\end{align}
	where $\mathbb{K}$ represents the nonlocal electromagnetic response kernel.
	In the microscopic theory, this quantity is expressed as the charge current-charge current correlation function at temperature $T$.
	Performing Fourier transformation over space and time gives the correlation function in momentum and frequency spaces as 
	%
	\begin{align}
		\begin{aligned}\bJ(\bq,\omega,T)=-\mathbb{K}(\bq,\omega,T)\bA(\bq,\omega).    \end{aligned}
	\end{align}
	%
	In the context of linear response theory and through the application of  the Kubo formula, this correlation function is defined  as
	%
	\begin{align}
		\begin{aligned}
			\mathbb{K}&^{}_{\alpha\beta}(\bq,\omega^{}_m,T)
			=
			\mathbb{K}^{}_{0}\delta^{}_{\alpha\beta}
			+
			\Delta \mathbb{K}^{}_{\alpha\beta}(\bq,{\mi}\omega,T)
			\Big\vert^{}_{{\mi}\omega\rightarrow \omega^{}_m+{\mi}0^+},
		\end{aligned}
		\label{Eq:Kubo_Formula}
	\end{align}
	%
	where
	$\mathbb{K}^{}_{0}={ne^2}/{m}$, and $n$, $e$ and $m$ are density,  charge and bare mass of electron, respectively. 
	It should be emphasized that $\mathbb{K}^{}_{0}$ is solely a band structure property and is related to the magnetic penetration depth $\lambda(0)=\lambda^{}_0$ at absolute zero temperature.
	Moreover, the temperature dependent part of the response function determining the deviation of magnetic penetration depth from $\lambda^{}_{0}$  is given by~\cite{Coleman_2015}
	\begin{align}
		\begin{aligned}
			\Delta \mathbb{K}^{}_{\alpha\beta}&(\bq,{\mi}\omega^{}_m,T)
			=
			\\
			&-\frac{T}{4N}
			\sum^{}_{\bk,{\mi}\nu^{}_n}
			\hat{J}^{\alpha}_{\bk}
			\hat{{\cal G}}^{}_{\bk}({\mi}\nu^{}_n)
			\hat{J}^{\beta}_{\bk}
			\hat{{\cal G}}^{}_{\bk+\bq}({\mi}\omega_m+{\mi}\nu^{}_n),
		\end{aligned}
		\label{Eq:Delta_K}
	\end{align}
	%
	in which $\omega^{}_{m}=2m\pi$ stands for bosonic Matsubara frequency.
	In addition,
	%
	\begin{equation}
		\hat{J}^{\alpha}_{\bk}
		=
		\begin{bmatrix}
			J^{\alpha,e}_{\bk} & 0
			\\
			0 & J^{\alpha,h}_{\bk} 
		\end{bmatrix},
		\label{Eq:Charge_Current_Matrix}
	\end{equation}
	%
	represents the $2\times 2$ matrix of the $\alpha^{\rm th}$ component of charge current in Nambu space $(a^{\dagger}_{\bk,-,\uparrow},a^{}_{-\bk,-,\downarrow})$, whose elements are defined by~\cite{Kontani_PRL_2009,Biderang_PhyisicaC_2016}
	\begin{align}
		\begin{aligned}
			J^{\alpha,e}_{\bk}=
			-\frac{\partial \xi^{}_{\bk+e\bA}}{\partial A^{}_{\alpha}}
			\Big\vert^{}_{\bA=0};\quad
			J^{\alpha,h}_{\bk}=
			\frac{\partial \xi^{}_{-\bk+e\bA}}{\partial A^{}_{\alpha}}
			\Big\vert^{}_{\bA=0}.
		\end{aligned}
		\label{Eq:J_k_e_k}
	\end{align}
	%
	%
	Furthermore, the Matsubara Green's function is 
	%
	\begin{equation} 
		\hat{G}(\bk,{\mi}\nu)=
		\begin{bmatrix}
			{\cal G}(\bk,{\mi}\nu^{}_{n}) && {\cal F}(\bk,{\mi}\nu^{}_{n})
			\\
			{\cal F}^{\dagger}(\bk,{\mi}\nu^{}_{n}) && -{\cal G}^{T}(-\bk,{-\mi} \nu^{}_{n})
		\end{bmatrix},
		\label{Eq:Sc_Green_Func}
	\end{equation}
	%
	where
	%
	\begin{align} 
		\begin{aligned}
			{\cal G}(\bk,{\mi}\nu^{}_{n}) 
			&= 
			\frac{ {\mi}\nu^{}_{n}  +  \xi_{\bk,-}}
			{({\mi}\nu^{}_{n})^2-E^2_{\bk,-}};
			\;\;
			{\cal F}(\bk,{\mi}\nu^{}_{n}) 
			= \frac{ -\Delta_{\bk}}  {({\mi}\nu^{}_{n})^2-E^2_{\bk,-}}
		\end{aligned}
		\label{eq:G_helicity}
	\end{align}
	%
	are normal and anomalous Green's functions, respectively, with the fermionic Matsubara  frequency $\nu_n=(2n+1)\pi T$ at finite  temperature $T$, and the quasi-particle  dispersion  
	$E_{\bk,-} = \sqrt{{\xi^2_{\bk,-} + \Delta^2_{\bk}}}$.
	Thus in the superconducting state, $\Delta \mathbb{K}^{}_{\alpha\beta}(\bq,\omega,T)$ can be simplified as
	%
	\begin{align}
		\begin{aligned}
			\Delta &\mathbb{K}^{}_{\alpha\beta}(\bq,\omega,T) 
			=
			\\
			&\frac{1}{N}	
			\sum_{\bk}
			J^{\alpha,e}_{\bk}J^{\beta,e}_{\bk}  
			\Big[
			\frac{n_f(E_{\bk,-}) - n_f(E_{\bk+\bq,-})}
			{\omega+E_{\bk,-}-E_{\bk+\bq,-}+{\mi}0^+}
			\Big],
		\end{aligned}
		\label{Eq:del_K_T}  
	\end{align}
	%
	where $n_f(\epsilon)=1/(1+\exp(\epsilon/T))$ is the Fermi-Dirac distribution function.
	Since we are specifically interested in the DC  response ($\omega=0$), afterwards we drop  $\omega$ from our calculations.
	In the local limit ($\bq\rightarrow0$), one finds
	%
	\begin{align}
		\begin{aligned}
			{\rm Re}
			\Big[\Delta \mathbb{K}^{}_{\alpha\beta}(\bq=0,T)
			\Big] 
			\!=\! 
			-\sum_{\bk} 
			\frac{J^{\alpha,e}_{\bk}J^{\beta,e}_{\bk}}
			{4T\cosh^{2}
				(E^{}_{\bk,-}/2T)}.
		\end{aligned}
		\label{Eq:K_final}
	\end{align}
	%
	While the magnetic penetration depth is  linked to the real component  of electromagnetic response tensor, we have~\cite{Tinkham_2004}
	%
	\begin{align}
		\lambda^{\rm loc}_{\alpha\beta}(T)=
		\Big[
		\frac{4\pi}{c}
		{\rm Re} [\mathbb{K}^{}_{\alpha\beta}(\bq=0,T)]
		\Big]^{-\frac{1}{2}}_{}.
		\label{Eq:Local_pen}
	\end{align}
	%
	In this context,  we can thus formulate  the magnetic penetration depth  for a superconductor with an arbitrary Fermi surface as 
	%
	\begin{align}
		\begin{aligned}
			\lambda^{\rm loc}_{\alpha\beta}(T) 
			= 
			\lambda_0\Bigg[
			\delta^{}_{\alpha\beta} 
			-
			\sum_{\bk} 
			\frac{J^{\alpha,e}_{\bk}J^{\beta,e}_{\bk} }{4\mathbb{K}^{}_{0} T} 
			\cosh^{-2}\left(\frac{E_{\bk}}{2T}\right)
			\Bigg]^{-\frac{1}{2}}.
		\end{aligned}
		\label{Eq:lambda_0_lambda_T}
	\end{align}
	%
	%
	Hence, within the low-temperature regime,  $T\ll T_C$, we can derive 	%
	\begin{equation}
		\lambda_{\rm loc}(T)
		\approx
		\lambda_0
		\left[
		\delta^{}_{\alpha\beta}
		+
		\sum_{\bk}
		\frac{J^{\alpha,e}_{\bk}J^{\beta,e}_{\bk}}{8 \mathbb{K}^{}_{0} T}
		\cosh^{-2}\left(\frac{E_{\bk}}{2T}\right)\right].
		\label{Eq:lambda_local_low_T}
	\end{equation}
	%
	%
	%
	However, for the nonlocal limit ($\bq\neq 0$), the electrons are reflected by the boundary either specularly or diffusively.
	In both regimes, the  penetration depth is related to the superconducting properties of the material, such as its critical temperature and the density of superconducting carriers. It is also influenced by external factors such as temperature and the frequency of the incident electromagnetic waves.
	\\	

	For the {\it specular boundary condition}, the magnetic penetration depth is determined by~\cite{Tinkham_2004}
	%
	\begin{align}
		\lambda^{}_{\rm spec}(T)=
		\frac{2}{\pi}
		\int^{\infty}_{0}
		dq
		\Big[
		q^2
		+
		{\rm Re}[\mathbb{K}(\bq,T)]
		\Big]^{-1}_{},
		\label{Eq:NonLocal_Spec}
	\end{align}
	where electromagnetic waves incident on the superconductor are reflected in a predictable and orderly manner, similar to the way light reflects off a mirror. This regime typically applies to smooth and clean surfaces of the superconductor. Moreover, the penetration depth refers to the depth at which the amplitude of the electromagnetic field decreases by a factor of $1/e$ (about $37\%$) compared to its value at the surface.
	\\

	On the other hand, when considering a {\it diffusive boundary}, one can find~\cite{Tinkham_2004}
	%
	%
	\begin{align}
		\lambda^{}_{\rm diff}(T)=
		\pi
		\Big[
		\int^{\infty}_{0}
		dq
		\ln
		\Big(
		1
		+
		\frac{4\pi}{c}
		\frac{{\rm Re}[\mathbb{K}(\bq,T)]}{q^2}
		\Big)
		\Big]^{-1}_{},
		\label{Eq:NonLocal_Diff}
	\end{align}
	%
	where 
	the electromagnetic waves incident on the superconductor are scattered in various directions due to surface roughness, impurities, or defects present on the surface. The  penetration depth   refers to the average distance over which the electromagnetic field is attenuated inside the superconductor. This regime typically applies to rough or dirty surfaces.
	\\
	
	We are also interested in the normalized superfluid density tensor, which can be expressed in terms of the magnetic penetration depth as~\cite{Sigrist_PRB_2006}
	%
	\begin{equation}
		\rho^s_{}(T) 
		=  
		\Big[    
		\frac{\lambda(T)}{\lambda_0}
		\Big]^{-2}.
		\label{Eq:Superfluid_Density}
	\end{equation}
	%
	For the simplistic case, namely an s-wave superconductor with a spherical Fermi surface  in the local limit, 
	our approach yields  
	%
	\begin{align}
		\begin{aligned}
			\rho^s(T) &=   
			1 + 2 
			\int d\varepsilon \frac{\partial{n_f(E_{\bk})}}{\partial E_{\bk}} 
			\\
			&=
			1 -\frac{1}{2T} 
			\int d\varepsilon 
			\cosh^{-2}
			\left(
			\frac{\sqrt{\varepsilon^2 + \Delta^2}}{2T}
			\right),
		\end{aligned}
		\label{Eq:Superfluid_Density_Special}
	\end{align}
	%
	which is the familiar expression of superfluid density for isotropic s-wave pairing \cite{Bauer_PRL_2005,Prozorov_2006}.

	This approach establishes the connection between the complex microscopic state for the most general superconductor and the macroscopic 
	penetration depth which can be experimentally  measured.
	Thus it can reveal the complex nature of superconducting gap and its pairing symmetry.
	In particular, an exponential or temperature-independent behavior  
	suggests a conventional s-wave superconducting gap (nodless).
	Conversely, at low temperatures, for unconventional pairing  the  penetration depth  can be approximated by   a power-law-type behavior, $\lambda(T) \propto T^n$, which can imply the existence of  nodal superconductivity.
	The type of the nodes in the system determines the  value of  the power exponent.
	Linear to quadratic temperature dependence   analogizes
	the presence of line nodes or point nodes in the superconducting gap.

	\begin{figure}[t]
		\includegraphics[width=\linewidth]{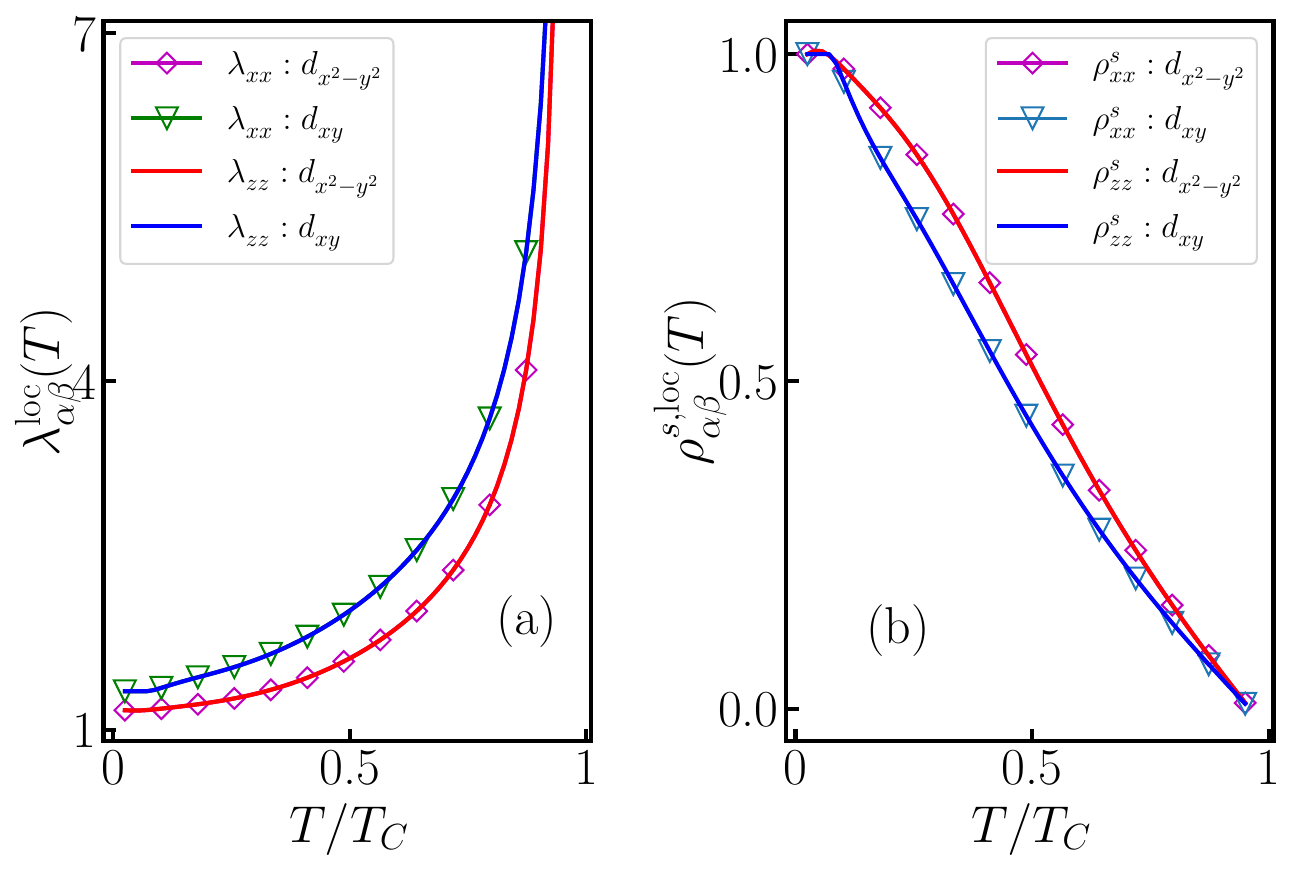}
		\caption{Temperature dependence of (a): $xx$- and $zz$-components of  penetration depth, and (d) superfluid density for possible $d^{}_{x^2-y^2}$- and $d^{}_{xy}$-wave pairings in CeCoIn$^{}_{5}$ in the local limit ($\bq\rightarrow0$).}
		\label{Fig:Local_London_SFD}
	\end{figure}

	\begin{figure}[t]
		\includegraphics[width=\linewidth]{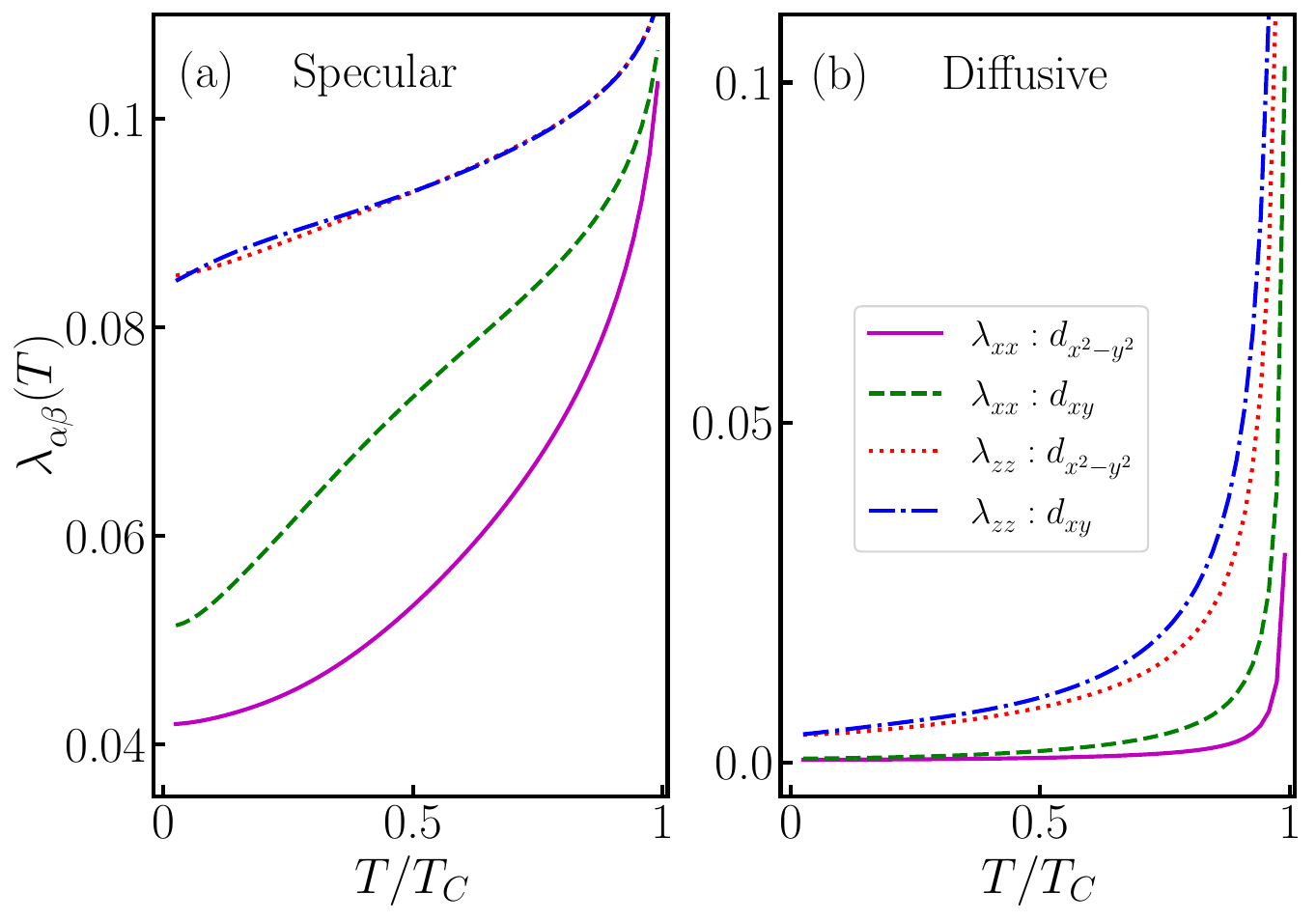}   %
		\caption{The  $xx$- and $zz$-components of $\lambda(T)$ for CeCoIn$_5$ in  (a): specular, and (b) diffusive regimes, considering possible $d^{}_{x^2-y^2}$- and $d^{}_{xy}$-wave  pairings.}
		\label{Fig:Non_Local_London}
	\end{figure}
	%

	\section{Results of penetration depth for ${\rm CeCoIn}_5$}
	In this section, we delve into the outcomes of our investigation into the penetration depth of CeCoIn$_5$, with a specific focus on symmetry-related aspects.
	It is important to note that due to the in-plane $C^{}_{4v}$ rotational symmetry, an expected equivalence arises between the $xx$- and $yy$-components of the  penetration depth and superfluid density tensors. 
	However, we should underscore that no such guarantee exists for the $zz$-component.
		Here in addition to the primary candidate pairing, $d^{}_{x^2-y^2}$, 
        $\Delta^{}_{0}(T)(\cos k_x-\cos k_y)$,
        we also investigate $d_{xy}$ pairing,  
        $\Delta^{}_{0}(T)\sin k_x \sin k_y$.
		The temperature dependence of superconducting order parameter is assumed to be of the form $\Delta^{}_{0}(T)=\Delta^{}_{0}\tanh[1.76\sqrt{T_C/T-1}]$, with $\Delta^{}_{0}$ as the amplitude of the pairing.
	%

	\begin{figure}[t]
		\includegraphics[width=\linewidth]{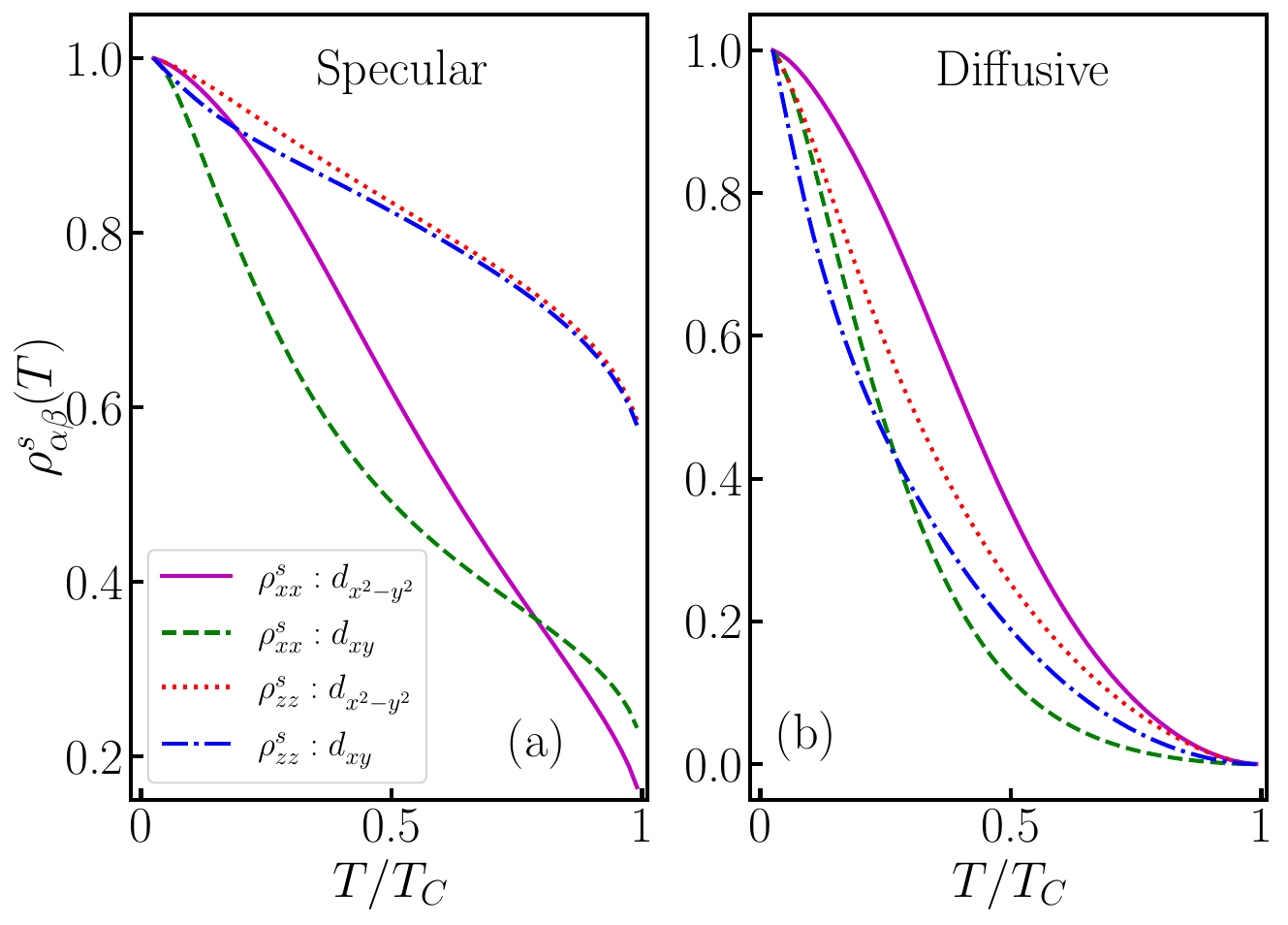}   %
		\caption{The spatial components of superfluid density tensor of possible $d_{x^2-y^2}$ and  $d_{xy}$ Cooper pairings in CeCoIn$_5$ for non-local (a) specular, and (b) diffusive regimes.}
		\label{Fig:Non_Local_SFD}
	\end{figure}
	%

	In the local limit (see Fig.~\ref{Fig:Local_London_SFD}), a theoretical prediction is fulfilled: all components of the  penetration depth and superfluid density tensors ($xx$, $yy$, and $zz$) converge to identical values. 
	Notably, within this limit, we made a significant observation the penetration depth of $d^{}_{xy}$-wave pairing surpasses that of $d^{}_{x^2-y^2}$-wave pairing. 
	Inversely, the superfluid density of $d^{}_{x^2-y^2}$ dominates over that of $d^{}_{xy}$.
	Moving into the non-local limit, Fig.~\ref{Fig:Non_Local_London}(a) investigates the specular scattering regime, revealing fascinating insights into the penetration depth's behavior.
	Particularly at higher temperatures, we observed a remarkable convergence in the $zz$-component values for both $d$-wave pairings, indicating a trend toward parity. 
	Notably, these $zz$-component values are significantly greater than those recorded for the $xx$- and $yy$-components. 
	Furthermore, we observe a striking dominance of the  penetration depth for $d^{}_{xy}$-wave pairing over its $d^{}_{x^2-y^2}$-wave counterpart in the $xx$- and $yy$-components. 
	In the diffusive regime, as shown in Fig.~\ref{Fig:Non_Local_London}(b), we have uncovered a distinct behavior within the  penetration depth characteristics. 
	In contrast to the specular regime, the $zz$-component exhibits a clear preference for $d^{}_{xy}$-wave pairing over $d^{}_{x^2-y^2}$-wave pairing. 
	This pattern is mirrored in the $xx$-components, further accentuating the dominance of $d^{}_{xy}$-wave pairing in this particular scattering regime. 
	Finally, Fig.~\ref{Fig:Non_Local_SFD} represents the temperature dependence of superfluid density in the non-local limit.
	In the specular regime, this quantity strongly depends on the symmetry of the superconducting gap.
	Besides, for every specific type of order parameter, an obvious difference is apparent between the $xx$- and $zz$- components of the superfluid density, specially at higher temperatures. 
	However, in the diffusive regime, there are lesser differences among the components of superfluid density for both $d^{}_{x^-y^2}$, and $d^{}_{xy}$ pairings. 
	\\

	
	\section{Summary}
	We present a novel microscopic formulation of the London (magnetic) penetration depth, specifically designed for heavy fermion superconductors, with a focus on CeCoIn$^{}_{5}$, a representative system known for its suggested $d$-wave pairings.
	Our formulation establishes a crucial link between the intricate microscopic superconducting state and the experimentally observable macroscopic properties in these materials.
	We demonstrate that the temperature-dependent behavior of the penetration depth is influenced by both the superconducting gap structure and the complex Fermi surface topology.
    Particularly,  our study provides evidence that CeCoIn$_5$ exhibits London-type superconductivity, where the penetration depth significantly exceeds the coherency length~\cite{Abrikosov_1975}.
    Furthermore,  in the local limit, 
    the temperature-dependent behavior of the London penetration depth suggests that the prevailing superconducting pairing mechanism for CeCoIn$_5$ is the nodal $d^{}_{x^2-y^2}$ symmetry. This features are closely aligned with Magnetic Force Microscopy results~\cite{JeeHooon_Kim_2020}.
    This versatile approach is applicable to a wide range of physical systems, making it valuable for exploring superconducting behaviors in various materials.
\\

	
	\section*{Acknowledgments}
	We acknowledge helpful discussions with   Geunyong Kim,  Jinyoung Youn, and M.~N.~Najafi.  We are grateful to Yunkyu Bang and Ki-Seok Kim for their insightful comments.
    M.B. thanks H. Yavari for valuable conversations.
%
	
	\bibliography{References}

\begin{thebibliography}{53}%
\makeatletter
\providecommand \@ifxundefined [1]{%
 \@ifx{#1\undefined}
}%
\providecommand \@ifnum [1]{%
 \ifnum #1\expandafter \@firstoftwo
 \else \expandafter \@secondoftwo
 \fi
}%
\providecommand \@ifx [1]{%
 \ifx #1\expandafter \@firstoftwo
 \else \expandafter \@secondoftwo
 \fi
}%
\providecommand \natexlab [1]{#1}%
\providecommand \enquote  [1]{``#1''}%
\providecommand \bibnamefont  [1]{#1}%
\providecommand \bibfnamefont [1]{#1}%
\providecommand \citenamefont [1]{#1}%
\providecommand \href@noop [0]{\@secondoftwo}%
\providecommand \href [0]{\begingroup \@sanitize@url \@href}%
\providecommand \@href[1]{\@@startlink{#1}\@@href}%
\providecommand \@@href[1]{\endgroup#1\@@endlink}%
\providecommand \@sanitize@url [0]{\catcode `\\12\catcode `\$12\catcode
  `\&12\catcode `\#12\catcode `\^12\catcode `\_12\catcode `\%12\relax}%
\providecommand \@@startlink[1]{}%
\providecommand \@@endlink[0]{}%
\providecommand \url  [0]{\begingroup\@sanitize@url \@url }%
\providecommand \@url [1]{\endgroup\@href {#1}{\urlprefix }}%
\providecommand \urlprefix  [0]{URL }%
\providecommand \Eprint [0]{\href }%
\providecommand \doibase [0]{https://doi.org/}%
\providecommand \selectlanguage [0]{\@gobble}%
\providecommand \bibinfo  [0]{\@secondoftwo}%
\providecommand \bibfield  [0]{\@secondoftwo}%
\providecommand \translation [1]{[#1]}%
\providecommand \BibitemOpen [0]{}%
\providecommand \bibitemStop [0]{}%
\providecommand \bibitemNoStop [0]{.\EOS\space}%
\providecommand \EOS [0]{\spacefactor3000\relax}%
\providecommand \BibitemShut  [1]{\csname bibitem#1\endcsname}%
\let\auto@bib@innerbib\@empty
\bibitem [{\citenamefont {{Bednorz}}\ and\ \citenamefont
  {{Müller}}(1986)}]{Muller_Bednorz_1986}%
  \BibitemOpen
  \bibfield  {author} {\bibinfo {author} {\bibfnamefont {J.~G.}\ \bibnamefont
  {{Bednorz}}}\ and\ \bibinfo {author} {\bibfnamefont {K.~A.}\ \bibnamefont
  {{Müller}}},\ }\bibfield  {title} {\bibinfo {title} {Possible high
  ${\mathrm{t}}^{}_{c}$ superconductivity in the {$\mathrm{BaLaCuO}$} system},\
  }\href {https://doi.org/10.1007/BF01303701} {\bibfield  {journal} {\bibinfo
  {journal} {Zeitschrift fur Physik B Condensed Matter}\ }\textbf {\bibinfo
  {volume} {64}},\ \bibinfo {pages} {189–193} (\bibinfo {year}
  {1986})}\BibitemShut {NoStop}%
\bibitem [{\citenamefont {Monthoux}\ \emph {et~al.}(2007)\citenamefont
  {Monthoux}, \citenamefont {Pines},\ and\ \citenamefont
  {Lonzarich}}]{Monthoux_2007}%
  \BibitemOpen
  \bibfield  {author} {\bibinfo {author} {\bibfnamefont {P.}~\bibnamefont
  {Monthoux}}, \bibinfo {author} {\bibfnamefont {D.}~\bibnamefont {Pines}},\
  and\ \bibinfo {author} {\bibfnamefont {G.~G.}\ \bibnamefont {Lonzarich}},\
  }\bibfield  {title} {\bibinfo {title} {Superconductivity without phonons},\
  }\href {https://doi.org/10.1038/nature06480} {\bibfield  {journal} {\bibinfo
  {journal} {Nature}\ }\textbf {\bibinfo {volume} {450}},\ \bibinfo {pages}
  {1177–1183} (\bibinfo {year} {2007})}\BibitemShut {NoStop}%
\bibitem [{\citenamefont {Norman}(2011)}]{Norman_2011}%
  \BibitemOpen
  \bibfield  {author} {\bibinfo {author} {\bibfnamefont {M.~R.}\ \bibnamefont
  {Norman}},\ }\bibfield  {title} {\bibinfo {title} {The challenge of
  unconventional superconductivity},\ }\href
  {https://doi.org/10.1126/science.1200181} {\bibfield  {journal} {\bibinfo
  {journal} {Science (New York, N.Y.)}\ }\textbf {\bibinfo {volume} {332}},\
  \bibinfo {pages} {196—200} (\bibinfo {year} {2011})}\BibitemShut {NoStop}%
\bibitem [{\citenamefont {Scalapino}(2012)}]{Scalapino_RevModPhys_1012}%
  \BibitemOpen
  \bibfield  {author} {\bibinfo {author} {\bibfnamefont {D.~J.}\ \bibnamefont
  {Scalapino}},\ }\bibfield  {title} {\bibinfo {title} {A common thread: The
  pairing interaction for unconventional superconductors},\ }\href
  {https://doi.org/10.1103/RevModPhys.84.1383} {\bibfield  {journal} {\bibinfo
  {journal} {Rev. Mod. Phys.}\ }\textbf {\bibinfo {volume} {84}},\ \bibinfo
  {pages} {1383–1417} (\bibinfo {year} {2012})}\BibitemShut {NoStop}%
\bibitem [{\citenamefont {Mineev}\ and\ \citenamefont
  {Samokhin}(1999)}]{Mineev-Samokhin-1999}%
  \BibitemOpen
  \bibfield  {author} {\bibinfo {author} {\bibfnamefont {V.}~\bibnamefont
  {Mineev}}\ and\ \bibinfo {author} {\bibfnamefont {K.}~\bibnamefont
  {Samokhin}},\ }\href {https://api.semanticscholar.org/CorpusID:119000791}
  {\emph {\bibinfo {title} {Introduction to Unconventional
  Superconductivity}}}\ (\bibinfo  {publisher} {Gordon and Breach, Science
  Publisher},\ \bibinfo {year} {1999})\BibitemShut {NoStop}%
\bibitem [{\citenamefont {Prozorov}\ and\ \citenamefont
  {Giannetta}(2006)}]{Prozorov_2006}%
  \BibitemOpen
  \bibfield  {author} {\bibinfo {author} {\bibfnamefont {R.}~\bibnamefont
  {Prozorov}}\ and\ \bibinfo {author} {\bibfnamefont {R.~W.}\ \bibnamefont
  {Giannetta}},\ }\bibfield  {title} {\bibinfo {title} {Magnetic penetration
  depth in unconventional superconductors},\ }\href
  {https://doi.org/10.1088/0953-2048/19/8/R01} {\bibfield  {journal} {\bibinfo
  {journal} {Superconductor Science and Technology}\ }\textbf {\bibinfo
  {volume} {19}},\ \bibinfo {pages} {R41} (\bibinfo {year} {2006})}\BibitemShut
  {NoStop}%
\bibitem [{\citenamefont {Abrikosov}\ \emph {et~al.}(1975)\citenamefont
  {Abrikosov}, \citenamefont {Gorkov},\ and\ \citenamefont
  {Dzyaloshinski}}]{Abrikosov_1975}%
  \BibitemOpen
  \bibfield  {author} {\bibinfo {author} {\bibfnamefont {A.}~\bibnamefont
  {Abrikosov}}, \bibinfo {author} {\bibfnamefont {L.}~\bibnamefont {Gorkov}},\
  and\ \bibinfo {author} {\bibfnamefont {I.}~\bibnamefont {Dzyaloshinski}},\
  }\href@noop {} {\emph {\bibinfo {title} {Methods of Quantum Field Theory in
  Statistical Physics}}}\ (\bibinfo  {publisher} {Dover Publications},\
  \bibinfo {year} {1975})\BibitemShut {NoStop}%
\bibitem [{\citenamefont {Eliashberg}\ \emph {et~al.}(1991)\citenamefont
  {Eliashberg}, \citenamefont {Klimovitch},\ and\ \citenamefont
  {Rylyakov}}]{Eliashberg_1991}%
  \BibitemOpen
  \bibfield  {author} {\bibinfo {author} {\bibfnamefont {G.}~\bibnamefont
  {Eliashberg}}, \bibinfo {author} {\bibfnamefont {G.}~\bibnamefont
  {Klimovitch}},\ and\ \bibinfo {author} {\bibfnamefont {A.}~\bibnamefont
  {Rylyakov}},\ }\bibfield  {title} {\bibinfo {title} {On the temperature
  dependence of the london penetration depth in a superconductor},\ }\href
  {https://doi.org/10.1007/BF00618221} {\bibfield  {journal} {\bibinfo
  {journal} {Journal of Superconductivity}\ }\textbf {\bibinfo {volume} {4}},\
  \bibinfo {pages} {393–396} (\bibinfo {year} {1991})}\BibitemShut {NoStop}%
\bibitem [{\citenamefont {Tinkham}(2004)}]{Tinkham_2004}%
  \BibitemOpen
  \bibfield  {author} {\bibinfo {author} {\bibfnamefont {M.}~\bibnamefont
  {Tinkham}},\ }\href {http://www.worldcat.org/isbn/0486435032} {\emph
  {\bibinfo {title} {Introduction to Superconductivity}}},\ Vol.\ \bibinfo
  {volume} {2nd edition}\ (\bibinfo  {publisher} {Dover Publications},\
  \bibinfo {year} {2004})\BibitemShut {NoStop}%
\bibitem [{\citenamefont {Bonalde}\ \emph {et~al.}(2009)\citenamefont
  {Bonalde}, \citenamefont {Ribeiro}, \citenamefont {Brämer-Escamilla},
  \citenamefont {Mu},\ and\ \citenamefont {Wen}}]{Wen_PRB_2009}%
  \BibitemOpen
  \bibfield  {author} {\bibinfo {author} {\bibfnamefont {I.}~\bibnamefont
  {Bonalde}}, \bibinfo {author} {\bibfnamefont {R.~L.}\ \bibnamefont
  {Ribeiro}}, \bibinfo {author} {\bibfnamefont {W.}~\bibnamefont
  {Brämer-Escamilla}}, \bibinfo {author} {\bibfnamefont {G.}~\bibnamefont
  {Mu}},\ and\ \bibinfo {author} {\bibfnamefont {H.~H.}\ \bibnamefont {Wen}},\
  }\bibfield  {title} {\bibinfo {title} {Possible two-gap behavior in
  noncentrosymmetric superconductor
  {${\text{Mg}}_{10}{\text{Ir}}_{19}{\text{B}}_{16}$}: A penetration depth
  study},\ }\href {https://doi.org/10.1103/PhysRevB.79.052506} {\bibfield
  {journal} {\bibinfo  {journal} {Phys. Rev. B}\ }\textbf {\bibinfo {volume}
  {79}},\ \bibinfo {pages} {052506} (\bibinfo {year} {2009})}\BibitemShut
  {NoStop}%
\bibitem [{\citenamefont {Kogan}\ and\ \citenamefont
  {Prozorov}(2021)}]{Prozonov_PRB_2021}%
  \BibitemOpen
  \bibfield  {author} {\bibinfo {author} {\bibfnamefont {V.~G.}\ \bibnamefont
  {Kogan}}\ and\ \bibinfo {author} {\bibfnamefont {R.}~\bibnamefont
  {Prozorov}},\ }\bibfield  {title} {\bibinfo {title} {Temperature dependence
  of london penetration depth anisotropy in superconductors with anisotropic
  order parameters},\ }\href {https://doi.org/10.1103/PhysRevB.103.054502}
  {\bibfield  {journal} {\bibinfo  {journal} {Phys. Rev. B}\ }\textbf {\bibinfo
  {volume} {103}},\ \bibinfo {pages} {054502} (\bibinfo {year}
  {2021})}\BibitemShut {NoStop}%
\bibitem [{\citenamefont {Barford}\ and\ \citenamefont
  {Gunn}(1988)}]{Barford_1988}%
  \BibitemOpen
  \bibfield  {author} {\bibinfo {author} {\bibfnamefont {W.}~\bibnamefont
  {Barford}}\ and\ \bibinfo {author} {\bibfnamefont {J.}~\bibnamefont {Gunn}},\
  }\bibfield  {title} {\bibinfo {title} {The theory of the measurement of the
  london penetration depth in uniaxial type ii superconductors by muon spin
  rotation},\ }\href {https://doi.org/10.1016/0921-4534(88)90014-7} {\bibfield
  {journal} {\bibinfo  {journal} {Physica C: Superconductivity}\ }\textbf
  {\bibinfo {volume} {156}},\ \bibinfo {pages} {515–522} (\bibinfo {year}
  {1988})}\BibitemShut {NoStop}%
\bibitem [{\citenamefont {Pümpin}\ \emph {et~al.}(1990)\citenamefont
  {Pümpin}, \citenamefont {Keller}, \citenamefont {Kündig}, \citenamefont
  {Odermatt}, \citenamefont {{Savi\ifmmode \acute{c}\else ć\fi{}}},
  \citenamefont {Schneider}, \citenamefont {Simmler}, \citenamefont
  {Zimmermann}, \citenamefont {Kaldis}, \citenamefont {Rusiecki}, \citenamefont
  {Maeno},\ and\ \citenamefont {Rossel}}]{Rossel_PRB_1990}%
  \BibitemOpen
  \bibfield  {author} {\bibinfo {author} {\bibfnamefont {B.}~\bibnamefont
  {Pümpin}}, \bibinfo {author} {\bibfnamefont {H.}~\bibnamefont {Keller}},
  \bibinfo {author} {\bibfnamefont {W.}~\bibnamefont {Kündig}}, \bibinfo
  {author} {\bibfnamefont {W.}~\bibnamefont {Odermatt}}, \bibinfo {author}
  {\bibfnamefont {I.~M.}\ \bibnamefont {{Savi\ifmmode \acute{c}\else
  ć\fi{}}}}, \bibinfo {author} {\bibfnamefont {J.~W.}\ \bibnamefont
  {Schneider}}, \bibinfo {author} {\bibfnamefont {H.}~\bibnamefont {Simmler}},
  \bibinfo {author} {\bibfnamefont {P.}~\bibnamefont {Zimmermann}}, \bibinfo
  {author} {\bibfnamefont {E.}~\bibnamefont {Kaldis}}, \bibinfo {author}
  {\bibfnamefont {S.}~\bibnamefont {Rusiecki}}, \bibinfo {author}
  {\bibfnamefont {Y.}~\bibnamefont {Maeno}},\ and\ \bibinfo {author}
  {\bibfnamefont {C.}~\bibnamefont {Rossel}},\ }\bibfield  {title} {\bibinfo
  {title} {Muon-spin-rotation measurements of the london penetration depths in
  {${\mathrm{YBa}}_{2}$${\mathrm{Cu}}_{3}$${\mathrm{O}}_{6.97}$}},\ }\href
  {https://doi.org/10.1103/PhysRevB.42.8019} {\bibfield  {journal} {\bibinfo
  {journal} {Phys. Rev. B}\ }\textbf {\bibinfo {volume} {42}},\ \bibinfo
  {pages} {8019–8029} (\bibinfo {year} {1990})}\BibitemShut {NoStop}%
\bibitem [{\citenamefont {Roseman}\ and\ \citenamefont
  {Grütter}(2001)}]{Roseman_2001}%
  \BibitemOpen
  \bibfield  {author} {\bibinfo {author} {\bibfnamefont {M.}~\bibnamefont
  {Roseman}}\ and\ \bibinfo {author} {\bibfnamefont {P.}~\bibnamefont
  {Grütter}},\ }\bibfield  {title} {\bibinfo {title} {Estimating the magnetic
  penetration depth using constant-height magnetic force microscopy images of
  vortices},\ }\href {https://doi.org/10.1088/1367-2630/3/1/324} {\bibfield
  {journal} {\bibinfo  {journal} {New Journal of Physics}\ }\textbf {\bibinfo
  {volume} {3}},\ \bibinfo {pages} {24} (\bibinfo {year} {2001})}\BibitemShut
  {NoStop}%
\bibitem [{\citenamefont {Khasanov}\ \emph {et~al.}(2006)\citenamefont
  {Khasanov}, \citenamefont {Landau}, \citenamefont {Baines}, \citenamefont
  {{La Mattina}}, \citenamefont {Maisuradze}, \citenamefont {Togano},\ and\
  \citenamefont {Keller}}]{Keller_PRB_2006}%
  \BibitemOpen
  \bibfield  {author} {\bibinfo {author} {\bibfnamefont {R.}~\bibnamefont
  {Khasanov}}, \bibinfo {author} {\bibfnamefont {I.~L.}\ \bibnamefont
  {Landau}}, \bibinfo {author} {\bibfnamefont {C.}~\bibnamefont {Baines}},
  \bibinfo {author} {\bibfnamefont {F.}~\bibnamefont {{La Mattina}}}, \bibinfo
  {author} {\bibfnamefont {A.}~\bibnamefont {Maisuradze}}, \bibinfo {author}
  {\bibfnamefont {K.}~\bibnamefont {Togano}},\ and\ \bibinfo {author}
  {\bibfnamefont {H.}~\bibnamefont {Keller}},\ }\bibfield  {title} {\bibinfo
  {title} {Muon-spin-rotation measurements of the penetration depth in
  {${\mathrm{Li}}_{2}{\mathrm{Pd}}_{3}\mathrm{B}$}},\ }\href
  {https://doi.org/10.1103/PhysRevB.73.214528} {\bibfield  {journal} {\bibinfo
  {journal} {Phys. Rev. B}\ }\textbf {\bibinfo {volume} {73}},\ \bibinfo
  {pages} {214528} (\bibinfo {year} {2006})}\BibitemShut {NoStop}%
\bibitem [{\citenamefont {Kim}\ \emph {et~al.}(2012)\citenamefont {Kim},
  \citenamefont {Civale}, \citenamefont {Nazaretski}, \citenamefont
  {Haberkorn}, \citenamefont {Ronning}, \citenamefont {Sefat}, \citenamefont
  {Tajima}, \citenamefont {Moeckly}, \citenamefont {Thompson},\ and\
  \citenamefont {Movshovich}}]{JeeHoonKim_2012}%
  \BibitemOpen
  \bibfield  {author} {\bibinfo {author} {\bibfnamefont {J.}~\bibnamefont
  {Kim}}, \bibinfo {author} {\bibfnamefont {L.}~\bibnamefont {Civale}},
  \bibinfo {author} {\bibfnamefont {E.}~\bibnamefont {Nazaretski}}, \bibinfo
  {author} {\bibfnamefont {N.}~\bibnamefont {Haberkorn}}, \bibinfo {author}
  {\bibfnamefont {F.}~\bibnamefont {Ronning}}, \bibinfo {author} {\bibfnamefont
  {A.~S.}\ \bibnamefont {Sefat}}, \bibinfo {author} {\bibfnamefont
  {T.}~\bibnamefont {Tajima}}, \bibinfo {author} {\bibfnamefont {B.~H.}\
  \bibnamefont {Moeckly}}, \bibinfo {author} {\bibfnamefont {J.~D.}\
  \bibnamefont {Thompson}},\ and\ \bibinfo {author} {\bibfnamefont
  {R.}~\bibnamefont {Movshovich}},\ }\bibfield  {title} {\bibinfo {title}
  {Direct measurement of the magnetic penetration depth by magnetic force
  microscopy},\ }\href {https://doi.org/10.1088/0953-2048/25/11/112001}
  {\bibfield  {journal} {\bibinfo  {journal} {Superconductor Science and
  Technology}\ }\textbf {\bibinfo {volume} {25}},\ \bibinfo {pages} {112001}
  (\bibinfo {year} {2012})}\BibitemShut {NoStop}%
\bibitem [{\citenamefont {Eltschka}\ \emph {et~al.}(2015)\citenamefont
  {Eltschka}, \citenamefont {Jaeck}, \citenamefont {Assig}, \citenamefont
  {Kondrashov}, \citenamefont {Skvortsov}, \citenamefont {Etzkorn},
  \citenamefont {Ast},\ and\ \citenamefont {Kern}}]{Eltschka_2015}%
  \BibitemOpen
  \bibfield  {author} {\bibinfo {author} {\bibfnamefont {M.}~\bibnamefont
  {Eltschka}}, \bibinfo {author} {\bibfnamefont {B.}~\bibnamefont {Jaeck}},
  \bibinfo {author} {\bibfnamefont {M.}~\bibnamefont {Assig}}, \bibinfo
  {author} {\bibfnamefont {O.}~\bibnamefont {Kondrashov}}, \bibinfo {author}
  {\bibfnamefont {M.}~\bibnamefont {Skvortsov}}, \bibinfo {author}
  {\bibfnamefont {M.}~\bibnamefont {Etzkorn}}, \bibinfo {author} {\bibfnamefont
  {C.}~\bibnamefont {Ast}},\ and\ \bibinfo {author} {\bibfnamefont
  {K.}~\bibnamefont {Kern}},\ }\bibfield  {title} {\bibinfo {title}
  {Superconducting scanning tunneling microscopy tips in a magnetic field:
  Geometry-controlled order of the phase transition},\ }\href
  {https://doi.org/10.1063/1.4931359} {\bibfield  {journal} {\bibinfo
  {journal} {Applied Physics Letters}\ }\textbf {\bibinfo {volume} {107}},\
  \bibinfo {pages} {122601} (\bibinfo {year} {2015})}\BibitemShut {NoStop}%
\bibitem [{\citenamefont {Coffey}(2000)}]{Coffey_JAP_2002}%
  \BibitemOpen
  \bibfield  {author} {\bibinfo {author} {\bibfnamefont {M.~W.}\ \bibnamefont
  {Coffey}},\ }\bibfield  {title} {\bibinfo {title} {Analyzing mutual
  inductance measurements to determine the london penetration depth},\ }\href
  {https://doi.org/10.1063/1.373076} {\bibfield  {journal} {\bibinfo  {journal}
  {Journal of Applied Physics}\ }\textbf {\bibinfo {volume} {87}},\ \bibinfo
  {pages} {4344–4351} (\bibinfo {year} {2000})},\ \Eprint
  {https://arxiv.org/abs/https://pubs.aip.org/aip/jap/article-pdf/87/9/4344/10604074/4344\_1\_online.pdf}
  {https://pubs.aip.org/aip/jap/article-pdf/87/9/4344/10604074/4344\_1\_online.pdf}
  \BibitemShut {NoStop}%
\bibitem [{\citenamefont {Harris}\ \emph {et~al.}(2010)\citenamefont {Harris},
  \citenamefont {Johansson}, \citenamefont {Berkley}, \citenamefont {Johnson},
  \citenamefont {Lanting}, \citenamefont {Han}, \citenamefont {Bunyk},
  \citenamefont {Ladizinsky}, \citenamefont {Oh}, \citenamefont {Perminov},
  \citenamefont {Tolkacheva}, \citenamefont {Uchaikin}, \citenamefont
  {Chapple}, \citenamefont {Enderud}, \citenamefont {Rich}, \citenamefont
  {Thom}, \citenamefont {Wang}, \citenamefont {Wilson},\ and\ \citenamefont
  {Rose}}]{Harris_PRB_2012}%
  \BibitemOpen
  \bibfield  {author} {\bibinfo {author} {\bibfnamefont {R.}~\bibnamefont
  {Harris}}, \bibinfo {author} {\bibfnamefont {J.}~\bibnamefont {Johansson}},
  \bibinfo {author} {\bibfnamefont {A.~J.}\ \bibnamefont {Berkley}}, \bibinfo
  {author} {\bibfnamefont {M.~W.}\ \bibnamefont {Johnson}}, \bibinfo {author}
  {\bibfnamefont {T.}~\bibnamefont {Lanting}}, \bibinfo {author} {\bibfnamefont
  {S.}~\bibnamefont {Han}}, \bibinfo {author} {\bibfnamefont {P.}~\bibnamefont
  {Bunyk}}, \bibinfo {author} {\bibfnamefont {E.}~\bibnamefont {Ladizinsky}},
  \bibinfo {author} {\bibfnamefont {T.}~\bibnamefont {Oh}}, \bibinfo {author}
  {\bibfnamefont {I.}~\bibnamefont {Perminov}}, \bibinfo {author}
  {\bibfnamefont {E.}~\bibnamefont {Tolkacheva}}, \bibinfo {author}
  {\bibfnamefont {S.}~\bibnamefont {Uchaikin}}, \bibinfo {author}
  {\bibfnamefont {E.~M.}\ \bibnamefont {Chapple}}, \bibinfo {author}
  {\bibfnamefont {C.}~\bibnamefont {Enderud}}, \bibinfo {author} {\bibfnamefont
  {C.}~\bibnamefont {Rich}}, \bibinfo {author} {\bibfnamefont {M.}~\bibnamefont
  {Thom}}, \bibinfo {author} {\bibfnamefont {J.}~\bibnamefont {Wang}}, \bibinfo
  {author} {\bibfnamefont {B.}~\bibnamefont {Wilson}},\ and\ \bibinfo {author}
  {\bibfnamefont {G.}~\bibnamefont {Rose}},\ }\bibfield  {title} {\bibinfo
  {title} {Experimental demonstration of a robust and scalable flux qubit},\
  }\href {https://doi.org/10.1103/PhysRevB.81.134510} {\bibfield  {journal}
  {\bibinfo  {journal} {Phys. Rev. B}\ }\textbf {\bibinfo {volume} {81}},\
  \bibinfo {pages} {134510} (\bibinfo {year} {2010})}\BibitemShut {NoStop}%
\bibitem [{\citenamefont {Kosztin}\ and\ \citenamefont
  {Leggett}(1997)}]{Leggett_PRL_1997}%
  \BibitemOpen
  \bibfield  {author} {\bibinfo {author} {\bibfnamefont {I.}~\bibnamefont
  {Kosztin}}\ and\ \bibinfo {author} {\bibfnamefont {A.~J.}\ \bibnamefont
  {Leggett}},\ }\bibfield  {title} {\bibinfo {title} {Nonlocal effects on the
  magnetic penetration depth in d-wave superconductors},\ }\href
  {https://doi.org/10.1103/PhysRevLett.79.135} {\bibfield  {journal} {\bibinfo
  {journal} {Phys. Rev. Lett.}\ }\textbf {\bibinfo {volume} {79}},\ \bibinfo
  {pages} {135–138} (\bibinfo {year} {1997})}\BibitemShut {NoStop}%
\bibitem [{\citenamefont {Hayashi}\ \emph {et~al.}(2006)\citenamefont
  {Hayashi}, \citenamefont {Wakabayashi}, \citenamefont {Frigeri},\ and\
  \citenamefont {Sigrist}}]{Sigrist_PRB_2006}%
  \BibitemOpen
  \bibfield  {author} {\bibinfo {author} {\bibfnamefont {N.}~\bibnamefont
  {Hayashi}}, \bibinfo {author} {\bibfnamefont {K.}~\bibnamefont
  {Wakabayashi}}, \bibinfo {author} {\bibfnamefont {P.~A.}\ \bibnamefont
  {Frigeri}},\ and\ \bibinfo {author} {\bibfnamefont {M.}~\bibnamefont
  {Sigrist}},\ }\bibfield  {title} {\bibinfo {title} {Temperature dependence of
  the superfluid density in a noncentrosymmetric superconductor},\ }\href
  {https://doi.org/10.1103/PhysRevB.73.024504} {\bibfield  {journal} {\bibinfo
  {journal} {Phys. Rev. B}\ }\textbf {\bibinfo {volume} {73}},\ \bibinfo
  {pages} {024504} (\bibinfo {year} {2006})}\BibitemShut {NoStop}%
\bibitem [{\citenamefont {Chen}\ \emph {et~al.}(2013)\citenamefont {Chen},
  \citenamefont {Jiao}, \citenamefont {Zhang}, \citenamefont {Chen},
  \citenamefont {Yang}, \citenamefont {Nicklas}, \citenamefont {Steglich},\
  and\ \citenamefont {Yuan}}]{Chen_2013}%
  \BibitemOpen
  \bibfield  {author} {\bibinfo {author} {\bibfnamefont {J.}~\bibnamefont
  {Chen}}, \bibinfo {author} {\bibfnamefont {L.}~\bibnamefont {Jiao}}, \bibinfo
  {author} {\bibfnamefont {J.~L.}\ \bibnamefont {Zhang}}, \bibinfo {author}
  {\bibfnamefont {Y.}~\bibnamefont {Chen}}, \bibinfo {author} {\bibfnamefont
  {L.}~\bibnamefont {Yang}}, \bibinfo {author} {\bibfnamefont {M.}~\bibnamefont
  {Nicklas}}, \bibinfo {author} {\bibfnamefont {F.}~\bibnamefont {Steglich}},\
  and\ \bibinfo {author} {\bibfnamefont {H.~Q.}\ \bibnamefont {Yuan}},\
  }\bibfield  {title} {\bibinfo {title} {Evidence for two-gap superconductivity
  in the non-centrosymmetric compound {LaNiC$_2$}},\ }\href
  {https://doi.org/10.1088/1367-2630/15/5/053005} {\bibfield  {journal}
  {\bibinfo  {journal} {New Journal of Physics}\ }\textbf {\bibinfo {volume}
  {15}},\ \bibinfo {pages} {053005} (\bibinfo {year} {2013})}\BibitemShut
  {NoStop}%
\bibitem [{\citenamefont {von Rohr}\ \emph {et~al.}(2019)\citenamefont {von
  Rohr}, \citenamefont {Orain}, \citenamefont {Khasanov}, \citenamefont
  {Witteveen}, \citenamefont {Shermadini}, \citenamefont {Nikitin},
  \citenamefont {Chang}, \citenamefont {Wieteska}, \citenamefont {Pasupathy},
  \citenamefont {Hasan}, \citenamefont {Amato}, \citenamefont {Luetkens},
  \citenamefont {Uemura},\ and\ \citenamefont {Guguchia}}]{Guguchia_2019}%
  \BibitemOpen
  \bibfield  {author} {\bibinfo {author} {\bibfnamefont {F.~O.}\ \bibnamefont
  {von Rohr}}, \bibinfo {author} {\bibfnamefont {J.-C.}\ \bibnamefont {Orain}},
  \bibinfo {author} {\bibfnamefont {R.}~\bibnamefont {Khasanov}}, \bibinfo
  {author} {\bibfnamefont {C.}~\bibnamefont {Witteveen}}, \bibinfo {author}
  {\bibfnamefont {Z.}~\bibnamefont {Shermadini}}, \bibinfo {author}
  {\bibfnamefont {A.}~\bibnamefont {Nikitin}}, \bibinfo {author} {\bibfnamefont
  {J.}~\bibnamefont {Chang}}, \bibinfo {author} {\bibfnamefont {A.~R.}\
  \bibnamefont {Wieteska}}, \bibinfo {author} {\bibfnamefont {A.~N.}\
  \bibnamefont {Pasupathy}}, \bibinfo {author} {\bibfnamefont {M.~Z.}\
  \bibnamefont {Hasan}}, \bibinfo {author} {\bibfnamefont {A.}~\bibnamefont
  {Amato}}, \bibinfo {author} {\bibfnamefont {H.}~\bibnamefont {Luetkens}},
  \bibinfo {author} {\bibfnamefont {Y.~J.}\ \bibnamefont {Uemura}},\ and\
  \bibinfo {author} {\bibfnamefont {Z.}~\bibnamefont {Guguchia}},\ }\bibfield
  {title} {\bibinfo {title} {Unconventional scaling of the superfluid density
  with the critical temperature in transition metal dichalcogenides},\ }\href
  {https://doi.org/10.1126/sciadv.aav8465} {\bibfield  {journal} {\bibinfo
  {journal} {Science Advances}\ }\textbf {\bibinfo {volume} {5}},\ \bibinfo
  {pages} {eaav8465} (\bibinfo {year} {2019})}\BibitemShut {NoStop}%
\bibitem [{\citenamefont {Wu}\ \emph {et~al.}(2020)\citenamefont {Wu},
  \citenamefont {Pal}, \citenamefont {Hosur},\ and\ \citenamefont
  {Foster}}]{Chun_PRL_2020}%
  \BibitemOpen
  \bibfield  {author} {\bibinfo {author} {\bibfnamefont {T.~C.}\ \bibnamefont
  {Wu}}, \bibinfo {author} {\bibfnamefont {H.~K.}\ \bibnamefont {Pal}},
  \bibinfo {author} {\bibfnamefont {P.}~\bibnamefont {Hosur}},\ and\ \bibinfo
  {author} {\bibfnamefont {M.~S.}\ \bibnamefont {Foster}},\ }\bibfield  {title}
  {\bibinfo {title} {Power-law temperature dependence of the penetration depth
  in a topological superconductor due to surface states},\ }\href
  {https://doi.org/10.1103/PhysRevLett.124.067001} {\bibfield  {journal}
  {\bibinfo  {journal} {Phys. Rev. Lett.}\ }\textbf {\bibinfo {volume} {124}},\
  \bibinfo {pages} {067001} (\bibinfo {year} {2020})}\BibitemShut {NoStop}%
\bibitem [{\citenamefont {Collomb}\ \emph {et~al.}(2021)\citenamefont
  {Collomb}, \citenamefont {Bending}, \citenamefont {Koshelev}, \citenamefont
  {Smylie}, \citenamefont {Farrar}, \citenamefont {Bao}, \citenamefont {Chung},
  \citenamefont {Kanatzidis}, \citenamefont {Kwok},\ and\ \citenamefont
  {Welp}}]{Welp_PRL_2021}%
  \BibitemOpen
  \bibfield  {author} {\bibinfo {author} {\bibfnamefont {D.}~\bibnamefont
  {Collomb}}, \bibinfo {author} {\bibfnamefont {S.~J.}\ \bibnamefont
  {Bending}}, \bibinfo {author} {\bibfnamefont {A.~E.}\ \bibnamefont
  {Koshelev}}, \bibinfo {author} {\bibfnamefont {M.~P.}\ \bibnamefont
  {Smylie}}, \bibinfo {author} {\bibfnamefont {L.}~\bibnamefont {Farrar}},
  \bibinfo {author} {\bibfnamefont {J.-K.}\ \bibnamefont {Bao}}, \bibinfo
  {author} {\bibfnamefont {D.~Y.}\ \bibnamefont {Chung}}, \bibinfo {author}
  {\bibfnamefont {M.~G.}\ \bibnamefont {Kanatzidis}}, \bibinfo {author}
  {\bibfnamefont {W.-K.}\ \bibnamefont {Kwok}},\ and\ \bibinfo {author}
  {\bibfnamefont {U.}~\bibnamefont {Welp}},\ }\bibfield  {title} {\bibinfo
  {title} {Observing the suppression of superconductivity in
  {${\mathrm{RbEuFe}}_{4}{\mathrm{As}}_{4}$} by correlated magnetic
  fluctuations},\ }\href {https://doi.org/10.1103/PhysRevLett.126.157001}
  {\bibfield  {journal} {\bibinfo  {journal} {Phys. Rev. Lett.}\ }\textbf
  {\bibinfo {volume} {126}},\ \bibinfo {pages} {157001} (\bibinfo {year}
  {2021})}\BibitemShut {NoStop}%
\bibitem [{\citenamefont {Dzhumanov}\ \emph {et~al.}(2022)\citenamefont
  {Dzhumanov}, \citenamefont {Turmanova},\ and\ \citenamefont
  {Kurbanov}}]{Dzhumanov_2022}%
  \BibitemOpen
  \bibfield  {author} {\bibinfo {author} {\bibfnamefont {S.}~\bibnamefont
  {Dzhumanov}}, \bibinfo {author} {\bibfnamefont {U.}~\bibnamefont
  {Turmanova}},\ and\ \bibinfo {author} {\bibfnamefont {U.}~\bibnamefont
  {Kurbanov}},\ }\bibfield  {title} {\bibinfo {title} {Two distinctive
  temperature dependences of the london penetration depth in high-tc cuprate
  superconductors as support for the theory of bose-liquid superconductivity},\
  }\href {https://doi.org/10.1016/j.physleta.2022.128447} {\bibfield  {journal}
  {\bibinfo  {journal} {Physics Letters A}\ }\textbf {\bibinfo {volume}
  {452}},\ \bibinfo {pages} {128447} (\bibinfo {year} {2022})}\BibitemShut
  {NoStop}%
\bibitem [{\citenamefont {Fletcher}\ \emph {et~al.}(2007)\citenamefont
  {Fletcher}, \citenamefont {Carrington}, \citenamefont {Diener}, \citenamefont
  {Rodière}, \citenamefont {Brison}, \citenamefont {Prozorov}, \citenamefont
  {Olheiser},\ and\ \citenamefont {Giannetta}}]{Fletcher_2007}%
  \BibitemOpen
  \bibfield  {author} {\bibinfo {author} {\bibfnamefont {J.}~\bibnamefont
  {Fletcher}}, \bibinfo {author} {\bibfnamefont {A.}~\bibnamefont
  {Carrington}}, \bibinfo {author} {\bibfnamefont {P.}~\bibnamefont {Diener}},
  \bibinfo {author} {\bibfnamefont {P.}~\bibnamefont {Rodière}}, \bibinfo
  {author} {\bibfnamefont {J.}~\bibnamefont {Brison}}, \bibinfo {author}
  {\bibfnamefont {R.}~\bibnamefont {Prozorov}}, \bibinfo {author}
  {\bibfnamefont {T.}~\bibnamefont {Olheiser}},\ and\ \bibinfo {author}
  {\bibfnamefont {R.}~\bibnamefont {Giannetta}},\ }\bibfield  {title} {\bibinfo
  {title} {Penetration depth study of superconducting gap structure of
  {2H-NbSe$_2$}},\ }\href {https://doi.org/10.1103/PhysRevLett.98.057003}
  {\bibfield  {journal} {\bibinfo  {journal} {Physical review letters}\
  }\textbf {\bibinfo {volume} {98}},\ \bibinfo {pages} {057003} (\bibinfo
  {year} {2007})}\BibitemShut {NoStop}%
\bibitem [{\citenamefont {Kim}\ \emph {et~al.}(2015)\citenamefont {Kim},
  \citenamefont {Tanatar}, \citenamefont {Flint}, \citenamefont {Petrovic},
  \citenamefont {Hu}, \citenamefont {White}, \citenamefont {Lum}, \citenamefont
  {Maple},\ and\ \citenamefont {Prozorov}}]{Prozonov_PRL_2015}%
  \BibitemOpen
  \bibfield  {author} {\bibinfo {author} {\bibfnamefont {H.}~\bibnamefont
  {Kim}}, \bibinfo {author} {\bibfnamefont {M.~A.}\ \bibnamefont {Tanatar}},
  \bibinfo {author} {\bibfnamefont {R.}~\bibnamefont {Flint}}, \bibinfo
  {author} {\bibfnamefont {C.}~\bibnamefont {Petrovic}}, \bibinfo {author}
  {\bibfnamefont {R.}~\bibnamefont {Hu}}, \bibinfo {author} {\bibfnamefont
  {B.~D.}\ \bibnamefont {White}}, \bibinfo {author} {\bibfnamefont {I.~K.}\
  \bibnamefont {Lum}}, \bibinfo {author} {\bibfnamefont {M.~B.}\ \bibnamefont
  {Maple}},\ and\ \bibinfo {author} {\bibfnamefont {R.}~\bibnamefont
  {Prozorov}},\ }\bibfield  {title} {\bibinfo {title} {Nodal to nodeless
  superconducting energy-gap structure change concomitant with fermi-surface
  reconstruction in the heavy-fermion compound {${\mathrm{CeCoIn}}_{5}$}},\
  }\href {https://doi.org/10.1103/PhysRevLett.114.027003} {\bibfield  {journal}
  {\bibinfo  {journal} {Phys. Rev. Lett.}\ }\textbf {\bibinfo {volume} {114}},\
  \bibinfo {pages} {027003} (\bibinfo {year} {2015})}\BibitemShut {NoStop}%
\bibitem [{\citenamefont {Steglich}\ \emph {et~al.}(1979)\citenamefont
  {Steglich}, \citenamefont {Aarts}, \citenamefont {Bredl}, \citenamefont
  {Lieke}, \citenamefont {Meschede}, \citenamefont {Franz},\ and\ \citenamefont
  {Schäfer}}]{Steglich-1979}%
  \BibitemOpen
  \bibfield  {author} {\bibinfo {author} {\bibfnamefont {F.}~\bibnamefont
  {Steglich}}, \bibinfo {author} {\bibfnamefont {J.}~\bibnamefont {Aarts}},
  \bibinfo {author} {\bibfnamefont {C.~D.}\ \bibnamefont {Bredl}}, \bibinfo
  {author} {\bibfnamefont {W.}~\bibnamefont {Lieke}}, \bibinfo {author}
  {\bibfnamefont {D.}~\bibnamefont {Meschede}}, \bibinfo {author}
  {\bibfnamefont {W.}~\bibnamefont {Franz}},\ and\ \bibinfo {author}
  {\bibfnamefont {H.}~\bibnamefont {Schäfer}},\ }\bibfield  {title} {\bibinfo
  {title} {Superconductivity in the presence of strong pauli paramagnetism:
  {CeCu$_2$Si$_2$}},\ }\href {https://doi.org/10.1103/PhysRevLett.43.1892}
  {\bibfield  {journal} {\bibinfo  {journal} {Physical Review Letters}\
  }\textbf {\bibinfo {volume} {43}},\ \bibinfo {pages} {1892–1896} (\bibinfo
  {year} {1979})}\BibitemShut {NoStop}%
\bibitem [{\citenamefont {Hewson}(1993)}]{Hewson-1993}%
  \BibitemOpen
  \bibfield  {author} {\bibinfo {author} {\bibfnamefont {A.}~\bibnamefont
  {Hewson}},\ }\href {https://doi.org/10.1017/CBO9780511470752} {\emph
  {\bibinfo {title} {The Kondo Problem to Heavy Fermions}}},\ Cambridge Studies
  in Magnetism\ (\bibinfo  {publisher} {Cambridge University Press},\ \bibinfo
  {year} {1993})\BibitemShut {NoStop}%
\bibitem [{\citenamefont {Pfleiderer}(2009)}]{Pfleiderer_REvModPhys_2009}%
  \BibitemOpen
  \bibfield  {author} {\bibinfo {author} {\bibfnamefont {C.}~\bibnamefont
  {Pfleiderer}},\ }\bibfield  {title} {\bibinfo {title} {Superconducting phases
  of $f$-electron compounds},\ }\href
  {https://doi.org/10.1103/RevModPhys.81.1551} {\bibfield  {journal} {\bibinfo
  {journal} {Rev. Mod. Phys.}\ }\textbf {\bibinfo {volume} {81}},\ \bibinfo
  {pages} {1551–1624} (\bibinfo {year} {2009})}\BibitemShut {NoStop}%
\bibitem [{\citenamefont {Petrovic}\ \emph {et~al.}(2001)\citenamefont
  {Petrovic}, \citenamefont {Pagliuso}, \citenamefont {Hundley}, \citenamefont
  {Movshovich}, \citenamefont {Sarrao}, \citenamefont {Thompson}, \citenamefont
  {Fisk},\ and\ \citenamefont {Monthoux}}]{Petrovic_2001}%
  \BibitemOpen
  \bibfield  {author} {\bibinfo {author} {\bibfnamefont {C.}~\bibnamefont
  {Petrovic}}, \bibinfo {author} {\bibfnamefont {P.~G.}\ \bibnamefont
  {Pagliuso}}, \bibinfo {author} {\bibfnamefont {M.~F.}\ \bibnamefont
  {Hundley}}, \bibinfo {author} {\bibfnamefont {R.}~\bibnamefont {Movshovich}},
  \bibinfo {author} {\bibfnamefont {J.~L.}\ \bibnamefont {Sarrao}}, \bibinfo
  {author} {\bibfnamefont {J.~D.}\ \bibnamefont {Thompson}}, \bibinfo {author}
  {\bibfnamefont {Z.}~\bibnamefont {Fisk}},\ and\ \bibinfo {author}
  {\bibfnamefont {P.}~\bibnamefont {Monthoux}},\ }\bibfield  {title} {\bibinfo
  {title} {Heavy-fermion superconductivity in {CeCoIn$_5$} at 2.3 k},\ }\href
  {https://doi.org/10.1088/0953-8984/13/17/103} {\bibfield  {journal} {\bibinfo
   {journal} {Journal of Physics: Condensed Matter}\ }\textbf {\bibinfo
  {volume} {13}},\ \bibinfo {pages} {L337} (\bibinfo {year}
  {2001})}\BibitemShut {NoStop}%
\bibitem [{\citenamefont {Bianchi}\ \emph {et~al.}(2003)\citenamefont
  {Bianchi}, \citenamefont {Movshovich}, \citenamefont {Vekhter}, \citenamefont
  {Pagliuso},\ and\ \citenamefont {Sarrao}}]{Bianchi_PRL_2003}%
  \BibitemOpen
  \bibfield  {author} {\bibinfo {author} {\bibfnamefont {A.}~\bibnamefont
  {Bianchi}}, \bibinfo {author} {\bibfnamefont {R.}~\bibnamefont {Movshovich}},
  \bibinfo {author} {\bibfnamefont {I.}~\bibnamefont {Vekhter}}, \bibinfo
  {author} {\bibfnamefont {P.~G.}\ \bibnamefont {Pagliuso}},\ and\ \bibinfo
  {author} {\bibfnamefont {J.~L.}\ \bibnamefont {Sarrao}},\ }\bibfield  {title}
  {\bibinfo {title} {Avoided antiferromagnetic order and quantum critical point
  in
  $\mathrm{C}\mathrm{e}\mathrm{C}\mathrm{o}\mathrm{I}{\mathrm{n}}_{\mathrm{5}}$},\
  }\href {https://doi.org/10.1103/PhysRevLett.91.257001} {\bibfield  {journal}
  {\bibinfo  {journal} {Phys. Rev. Lett.}\ }\textbf {\bibinfo {volume} {91}},\
  \bibinfo {pages} {257001} (\bibinfo {year} {2003})}\BibitemShut {NoStop}%
\bibitem [{\citenamefont {Paglione}\ \emph {et~al.}(2003)\citenamefont
  {Paglione}, \citenamefont {Tanatar}, \citenamefont {Hawthorn}, \citenamefont
  {Boaknin}, \citenamefont {Hill}, \citenamefont {Ronning}, \citenamefont
  {Sutherland}, \citenamefont {Taillefer}, \citenamefont {Petrovic},\ and\
  \citenamefont {Canfield}}]{Paglione_PRL_2003}%
  \BibitemOpen
  \bibfield  {author} {\bibinfo {author} {\bibfnamefont {J.}~\bibnamefont
  {Paglione}}, \bibinfo {author} {\bibfnamefont {M.~A.}\ \bibnamefont
  {Tanatar}}, \bibinfo {author} {\bibfnamefont {D.~G.}\ \bibnamefont
  {Hawthorn}}, \bibinfo {author} {\bibfnamefont {E.}~\bibnamefont {Boaknin}},
  \bibinfo {author} {\bibfnamefont {R.~W.}\ \bibnamefont {Hill}}, \bibinfo
  {author} {\bibfnamefont {F.}~\bibnamefont {Ronning}}, \bibinfo {author}
  {\bibfnamefont {M.}~\bibnamefont {Sutherland}}, \bibinfo {author}
  {\bibfnamefont {L.}~\bibnamefont {Taillefer}}, \bibinfo {author}
  {\bibfnamefont {C.}~\bibnamefont {Petrovic}},\ and\ \bibinfo {author}
  {\bibfnamefont {P.~C.}\ \bibnamefont {Canfield}},\ }\bibfield  {title}
  {\bibinfo {title} {Field-induced quantum critical point in
  {${\mathrm{C}\mathrm{e}\mathrm{C}\mathrm{o}\mathrm{I}\mathrm{n}}_{5}$}},\
  }\href {https://doi.org/10.1103/PhysRevLett.91.246405} {\bibfield  {journal}
  {\bibinfo  {journal} {Phys. Rev. Lett.}\ }\textbf {\bibinfo {volume} {91}},\
  \bibinfo {pages} {246405} (\bibinfo {year} {2003})}\BibitemShut {NoStop}%
\bibitem [{\citenamefont {{L. Sarrao}}\ and\ \citenamefont {{D.
  Thompson}}(2007)}]{Sarrao_2007}%
  \BibitemOpen
  \bibfield  {author} {\bibinfo {author} {\bibfnamefont {J.}~\bibnamefont {{L.
  Sarrao}}}\ and\ \bibinfo {author} {\bibfnamefont {J.}~\bibnamefont {{D.
  Thompson}}},\ }\bibfield  {title} {\bibinfo {title} {Superconductivity in
  cerium- and plutonium-based `115' materials},\ }\href
  {https://doi.org/10.1143/JPSJ.76.051013} {\bibfield  {journal} {\bibinfo
  {journal} {Journal of the Physical Society of Japan}\ }\textbf {\bibinfo
  {volume} {76}},\ \bibinfo {pages} {051013} (\bibinfo {year}
  {2007})}\BibitemShut {NoStop}%
\bibitem [{\citenamefont {{D. Thompson}}\ and\ \citenamefont
  {Fisk}(2012)}]{Fisk_2012}%
  \BibitemOpen
  \bibfield  {author} {\bibinfo {author} {\bibfnamefont {J.}~\bibnamefont {{D.
  Thompson}}}\ and\ \bibinfo {author} {\bibfnamefont {Z.}~\bibnamefont
  {Fisk}},\ }\bibfield  {title} {\bibinfo {title} {Progress in heavy-fermion
  superconductivity: {Ce115} and related materials},\ }\href
  {https://doi.org/10.1143/JPSJ.81.011002} {\bibfield  {journal} {\bibinfo
  {journal} {Journal of the Physical Society of Japan}\ }\textbf {\bibinfo
  {volume} {81}},\ \bibinfo {pages} {011002} (\bibinfo {year}
  {2012})}\BibitemShut {NoStop}%
\bibitem [{\citenamefont {Gyenis}\ \emph {et~al.}(2018)\citenamefont {Gyenis},
  \citenamefont {Feldman}, \citenamefont {Randeria}, \citenamefont {Peterson},
  \citenamefont {Bauer}, \citenamefont {Aynajian},\ and\ \citenamefont
  {Yazdani}}]{AliYazdani_NatComm_2018}%
  \BibitemOpen
  \bibfield  {author} {\bibinfo {author} {\bibfnamefont {A.}~\bibnamefont
  {Gyenis}}, \bibinfo {author} {\bibfnamefont {B.}~\bibnamefont {Feldman}},
  \bibinfo {author} {\bibfnamefont {M.}~\bibnamefont {Randeria}}, \bibinfo
  {author} {\bibfnamefont {G.}~\bibnamefont {Peterson}}, \bibinfo {author}
  {\bibfnamefont {E.}~\bibnamefont {Bauer}}, \bibinfo {author} {\bibfnamefont
  {P.}~\bibnamefont {Aynajian}},\ and\ \bibinfo {author} {\bibfnamefont
  {A.}~\bibnamefont {Yazdani}},\ }\bibfield  {title} {\bibinfo {title}
  {Visualizing heavy fermion confinement and pauli-limited superconductivity in
  layered cecoin5},\ }\bibfield  {journal} {\bibinfo  {journal} {Nature
  Communications}\ }\textbf {\bibinfo {volume} {9}},\ \href
  {https://doi.org/10.1038/s41467-018-02841-9} {10.1038/s41467-018-02841-9}
  (\bibinfo {year} {2018})\BibitemShut {NoStop}%
\bibitem [{\citenamefont {Sidorov}\ \emph {et~al.}(2002)\citenamefont
  {Sidorov}, \citenamefont {Nicklas}, \citenamefont {Pagliuso}, \citenamefont
  {Sarrao}, \citenamefont {Bang}, \citenamefont {Balatsky},\ and\ \citenamefont
  {Thompson}}]{Sidorov_PRL_2002}%
  \BibitemOpen
  \bibfield  {author} {\bibinfo {author} {\bibfnamefont {V.~A.}\ \bibnamefont
  {Sidorov}}, \bibinfo {author} {\bibfnamefont {M.}~\bibnamefont {Nicklas}},
  \bibinfo {author} {\bibfnamefont {P.~G.}\ \bibnamefont {Pagliuso}}, \bibinfo
  {author} {\bibfnamefont {J.~L.}\ \bibnamefont {Sarrao}}, \bibinfo {author}
  {\bibfnamefont {Y.}~\bibnamefont {Bang}}, \bibinfo {author} {\bibfnamefont
  {A.~V.}\ \bibnamefont {Balatsky}},\ and\ \bibinfo {author} {\bibfnamefont
  {J.~D.}\ \bibnamefont {Thompson}},\ }\bibfield  {title} {\bibinfo {title}
  {Superconductivity and quantum criticality in
  $\mathrm{C}\mathrm{e}\mathrm{C}\mathrm{o}\mathrm{I}{\mathrm{n}}_{\mathrm{5}}$},\
  }\href {https://doi.org/10.1103/PhysRevLett.89.157004} {\bibfield  {journal}
  {\bibinfo  {journal} {Phys. Rev. Lett.}\ }\textbf {\bibinfo {volume} {89}},\
  \bibinfo {pages} {157004} (\bibinfo {year} {2002})}\BibitemShut {NoStop}%
\bibitem [{\citenamefont {Izawa}\ \emph {et~al.}(2001)\citenamefont {Izawa},
  \citenamefont {Yamaguchi}, \citenamefont {Matsuda}, \citenamefont {Shishido},
  \citenamefont {Settai},\ and\ \citenamefont {Onuki}}]{Izawa_PRL_2001}%
  \BibitemOpen
  \bibfield  {author} {\bibinfo {author} {\bibfnamefont {K.}~\bibnamefont
  {Izawa}}, \bibinfo {author} {\bibfnamefont {H.}~\bibnamefont {Yamaguchi}},
  \bibinfo {author} {\bibfnamefont {Y.}~\bibnamefont {Matsuda}}, \bibinfo
  {author} {\bibfnamefont {H.}~\bibnamefont {Shishido}}, \bibinfo {author}
  {\bibfnamefont {R.}~\bibnamefont {Settai}},\ and\ \bibinfo {author}
  {\bibfnamefont {Y.}~\bibnamefont {Onuki}},\ }\bibfield  {title} {\bibinfo
  {title} {Angular position of nodes in the superconducting gap of quasi-2d
  heavy-fermion superconductor {${\mathrm{CeCoIn}}_{5}$}},\ }\href
  {https://doi.org/10.1103/PhysRevLett.87.057002} {\bibfield  {journal}
  {\bibinfo  {journal} {Phys. Rev. Lett.}\ }\textbf {\bibinfo {volume} {87}},\
  \bibinfo {pages} {057002} (\bibinfo {year} {2001})}\BibitemShut {NoStop}%
\bibitem [{\citenamefont {Park}\ \emph {et~al.}(2008)\citenamefont {Park},
  \citenamefont {Sarrao}, \citenamefont {Thompson},\ and\ \citenamefont
  {Greene}}]{Greene_PRL_2008}%
  \BibitemOpen
  \bibfield  {author} {\bibinfo {author} {\bibfnamefont {W.~K.}\ \bibnamefont
  {Park}}, \bibinfo {author} {\bibfnamefont {J.~L.}\ \bibnamefont {Sarrao}},
  \bibinfo {author} {\bibfnamefont {J.~D.}\ \bibnamefont {Thompson}},\ and\
  \bibinfo {author} {\bibfnamefont {L.~H.}\ \bibnamefont {Greene}},\ }\bibfield
   {title} {\bibinfo {title} {Andreev reflection in heavy-fermion
  superconductors and order parameter symmetry in {${\mathrm{CeCoIn}}_{5}$}},\
  }\href {https://doi.org/10.1103/PhysRevLett.100.177001} {\bibfield  {journal}
  {\bibinfo  {journal} {Phys. Rev. Lett.}\ }\textbf {\bibinfo {volume} {100}},\
  \bibinfo {pages} {177001} (\bibinfo {year} {2008})}\BibitemShut {NoStop}%
\bibitem [{\citenamefont {Stock}\ \emph {et~al.}(2008)\citenamefont {Stock},
  \citenamefont {Broholm}, \citenamefont {Hudis}, \citenamefont {Kang},\ and\
  \citenamefont {Petrovic}}]{Stock_PRL_2008}%
  \BibitemOpen
  \bibfield  {author} {\bibinfo {author} {\bibfnamefont {C.}~\bibnamefont
  {Stock}}, \bibinfo {author} {\bibfnamefont {C.}~\bibnamefont {Broholm}},
  \bibinfo {author} {\bibfnamefont {J.}~\bibnamefont {Hudis}}, \bibinfo
  {author} {\bibfnamefont {H.~J.}\ \bibnamefont {Kang}},\ and\ \bibinfo
  {author} {\bibfnamefont {C.}~\bibnamefont {Petrovic}},\ }\bibfield  {title}
  {\bibinfo {title} {Spin resonance in the $d$-wave superconductor
  {${\mathrm{CeCoIn}}_{5}$}},\ }\href
  {https://doi.org/10.1103/PhysRevLett.100.087001} {\bibfield  {journal}
  {\bibinfo  {journal} {Phys. Rev. Lett.}\ }\textbf {\bibinfo {volume} {100}},\
  \bibinfo {pages} {087001} (\bibinfo {year} {2008})}\BibitemShut {NoStop}%
\bibitem [{\citenamefont {Eremin}\ \emph {et~al.}(2008)\citenamefont {Eremin},
  \citenamefont {Zwicknagl}, \citenamefont {Thalmeier},\ and\ \citenamefont
  {Fulde}}]{PhysRevLett.101.187001}%
  \BibitemOpen
  \bibfield  {author} {\bibinfo {author} {\bibfnamefont {I.}~\bibnamefont
  {Eremin}}, \bibinfo {author} {\bibfnamefont {G.}~\bibnamefont {Zwicknagl}},
  \bibinfo {author} {\bibfnamefont {P.}~\bibnamefont {Thalmeier}},\ and\
  \bibinfo {author} {\bibfnamefont {P.}~\bibnamefont {Fulde}},\ }\bibfield
  {title} {\bibinfo {title} {Feedback spin resonance in superconducting
  {${\mathrm{CeCu}}_{2}{\mathrm{Si}}_{2}$ and ${\mathrm{CeCoIn}}_{5}$}},\
  }\href {https://doi.org/10.1103/PhysRevLett.101.187001} {\bibfield  {journal}
  {\bibinfo  {journal} {Phys. Rev. Lett.}\ }\textbf {\bibinfo {volume} {101}},\
  \bibinfo {pages} {187001} (\bibinfo {year} {2008})}\BibitemShut {NoStop}%
\bibitem [{\citenamefont {Akbari}\ and\ \citenamefont
  {Thalmeier}(2012)}]{PhysRevB.86.134516}%
  \BibitemOpen
  \bibfield  {author} {\bibinfo {author} {\bibfnamefont {A.}~\bibnamefont
  {Akbari}}\ and\ \bibinfo {author} {\bibfnamefont {P.}~\bibnamefont
  {Thalmeier}},\ }\bibfield  {title} {\bibinfo {title} {Field-induced spin
  exciton doublet splitting in {${d}_{{x}^{2}\ensuremath{-}{y}^{2}}$}-wave
  {{Ce}M{{In}}$_{5}$(M={Rh},{Ir},{Co})} heavy-electron superconductors},\
  }\href {https://doi.org/10.1103/PhysRevB.86.134516} {\bibfield  {journal}
  {\bibinfo  {journal} {Phys. Rev. B}\ }\textbf {\bibinfo {volume} {86}},\
  \bibinfo {pages} {134516} (\bibinfo {year} {2012})}\BibitemShut {NoStop}%
\bibitem [{\citenamefont {An}\ \emph {et~al.}(2010)\citenamefont {An},
  \citenamefont {Sakakibara}, \citenamefont {Settai}, \citenamefont {Onuki},
  \citenamefont {Hiragi}, \citenamefont {Ichioka},\ and\ \citenamefont
  {Machida}}]{PhysRevLett.104.037002}%
  \BibitemOpen
  \bibfield  {author} {\bibinfo {author} {\bibfnamefont {K.}~\bibnamefont
  {An}}, \bibinfo {author} {\bibfnamefont {T.}~\bibnamefont {Sakakibara}},
  \bibinfo {author} {\bibfnamefont {R.}~\bibnamefont {Settai}}, \bibinfo
  {author} {\bibfnamefont {Y.}~\bibnamefont {Onuki}}, \bibinfo {author}
  {\bibfnamefont {M.}~\bibnamefont {Hiragi}}, \bibinfo {author} {\bibfnamefont
  {M.}~\bibnamefont {Ichioka}},\ and\ \bibinfo {author} {\bibfnamefont
  {K.}~\bibnamefont {Machida}},\ }\bibfield  {title} {\bibinfo {title} {Sign
  reversal of field-angle resolved heat capacity oscillations in a heavy
  fermion superconductor ${\mathrm{cecoin}}_{5}$ and
  ${d}_{{x}^{2}\ensuremath{-}{y}^{2}}$ pairing symmetry},\ }\href
  {https://doi.org/10.1103/PhysRevLett.104.037002} {\bibfield  {journal}
  {\bibinfo  {journal} {Phys. Rev. Lett.}\ }\textbf {\bibinfo {volume} {104}},\
  \bibinfo {pages} {037002} (\bibinfo {year} {2010})}\BibitemShut {NoStop}%
\bibitem [{\citenamefont {Akbari}\ \emph {et~al.}(2011)\citenamefont {Akbari},
  \citenamefont {Thalmeier},\ and\ \citenamefont {Eremin}}]{Akbari_PRB_2011}%
  \BibitemOpen
  \bibfield  {author} {\bibinfo {author} {\bibfnamefont {A.}~\bibnamefont
  {Akbari}}, \bibinfo {author} {\bibfnamefont {P.}~\bibnamefont {Thalmeier}},\
  and\ \bibinfo {author} {\bibfnamefont {I.}~\bibnamefont {Eremin}},\
  }\bibfield  {title} {\bibinfo {title} {Quasiparticle interference in the
  heavy-fermion superconductor {${\mathrm{CeCoIn}}_{5}$}},\ }\href
  {https://doi.org/10.1103/PhysRevB.84.134505} {\bibfield  {journal} {\bibinfo
  {journal} {Phys. Rev. B}\ }\textbf {\bibinfo {volume} {84}},\ \bibinfo
  {pages} {134505} (\bibinfo {year} {2011})}\BibitemShut {NoStop}%
\bibitem [{\citenamefont {Allan}\ \emph {et~al.}(2013)\citenamefont {Allan},
  \citenamefont {Massee}, \citenamefont {Morr}, \citenamefont {Dyke},
  \citenamefont {Rost}, \citenamefont {Mackenzie}, \citenamefont {Petrovic},\
  and\ \citenamefont {Davis}}]{Allan_2013}%
  \BibitemOpen
  \bibfield  {author} {\bibinfo {author} {\bibfnamefont {M.}~\bibnamefont
  {Allan}}, \bibinfo {author} {\bibfnamefont {F.}~\bibnamefont {Massee}},
  \bibinfo {author} {\bibfnamefont {D.}~\bibnamefont {Morr}}, \bibinfo {author}
  {\bibfnamefont {J.}~\bibnamefont {Dyke}}, \bibinfo {author} {\bibfnamefont
  {A.}~\bibnamefont {Rost}}, \bibinfo {author} {\bibfnamefont {A.}~\bibnamefont
  {Mackenzie}}, \bibinfo {author} {\bibfnamefont {C.}~\bibnamefont
  {Petrovic}},\ and\ \bibinfo {author} {\bibfnamefont {J.}~\bibnamefont
  {Davis}},\ }\bibfield  {title} {\bibinfo {title} {Imaging cooper pairing of
  heavy fermions in {CeCoIn$_5$}},\ }\bibfield  {journal} {\bibinfo  {journal}
  {Nature Physics}\ }\textbf {\bibinfo {volume} {9}},\ \href
  {https://doi.org/10.1038/nphys2671} {10.1038/nphys2671} (\bibinfo {year}
  {2013})\BibitemShut {NoStop}%
\bibitem [{\citenamefont {Zhou}\ \emph {et~al.}(2013)\citenamefont {Zhou},
  \citenamefont {Misra}, \citenamefont {Neto}, \citenamefont {Aynajian},
  \citenamefont {Baumbach}, \citenamefont {Thompson}, \citenamefont {Bauer},\
  and\ \citenamefont {Yazdani}}]{AliYazdani_Nature_2013}%
  \BibitemOpen
  \bibfield  {author} {\bibinfo {author} {\bibfnamefont {B.}~\bibnamefont
  {Zhou}}, \bibinfo {author} {\bibfnamefont {S.}~\bibnamefont {Misra}},
  \bibinfo {author} {\bibfnamefont {E.}~\bibnamefont {Neto}}, \bibinfo {author}
  {\bibfnamefont {P.}~\bibnamefont {Aynajian}}, \bibinfo {author}
  {\bibfnamefont {R.}~\bibnamefont {Baumbach}}, \bibinfo {author}
  {\bibfnamefont {J.}~\bibnamefont {Thompson}}, \bibinfo {author}
  {\bibfnamefont {E.}~\bibnamefont {Bauer}},\ and\ \bibinfo {author}
  {\bibfnamefont {A.}~\bibnamefont {Yazdani}},\ }\bibfield  {title} {\bibinfo
  {title} {Visualizing nodal heavy fermion superconductivity in {CeCoIn$_5$}},\
  }\bibfield  {journal} {\bibinfo  {journal} {Nature Physics}\ }\textbf
  {\bibinfo {volume} {9}},\ \href {https://doi.org/10.1038/nphys2672}
  {10.1038/nphys2672} (\bibinfo {year} {2013})\BibitemShut {NoStop}%
\bibitem [{\citenamefont {Wulferding}\ \emph {et~al.}(2020)\citenamefont
  {Wulferding}, \citenamefont {Kim}, \citenamefont {Kim}, \citenamefont {Yang},
  \citenamefont {Bauer}, \citenamefont {Ronning}, \citenamefont {Movshovich},\
  and\ \citenamefont {Kim}}]{JeeHooon_Kim_2020}%
  \BibitemOpen
  \bibfield  {author} {\bibinfo {author} {\bibfnamefont {D.}~\bibnamefont
  {Wulferding}}, \bibinfo {author} {\bibfnamefont {G.}~\bibnamefont {Kim}},
  \bibinfo {author} {\bibfnamefont {H.}~\bibnamefont {Kim}}, \bibinfo {author}
  {\bibfnamefont {I.}~\bibnamefont {Yang}}, \bibinfo {author} {\bibfnamefont
  {E.~D.}\ \bibnamefont {Bauer}}, \bibinfo {author} {\bibfnamefont
  {F.}~\bibnamefont {Ronning}}, \bibinfo {author} {\bibfnamefont
  {R.}~\bibnamefont {Movshovich}},\ and\ \bibinfo {author} {\bibfnamefont
  {J.}~\bibnamefont {Kim}},\ }\bibfield  {title} {\bibinfo {title} {Local
  characterization of a heavy-fermion superconductor via sub-kelvin magnetic
  force microscopy},\ }\href {https://doi.org/10.1063/5.0028517} {\bibfield
  {journal} {\bibinfo  {journal} {Applied Physics Letters}\ }\textbf {\bibinfo
  {volume} {117}},\ \bibinfo {pages} {252601} (\bibinfo {year}
  {2020})}\BibitemShut {NoStop}%
\bibitem [{\citenamefont {Tanaka}\ \emph {et~al.}(2006)\citenamefont {Tanaka},
  \citenamefont {Ikeda}, \citenamefont {Nisikawa},\ and\ \citenamefont
  {Yamada}}]{Tanaka_JKPS_2006}%
  \BibitemOpen
  \bibfield  {author} {\bibinfo {author} {\bibfnamefont {K.}~\bibnamefont
  {Tanaka}}, \bibinfo {author} {\bibfnamefont {H.}~\bibnamefont {Ikeda}},
  \bibinfo {author} {\bibfnamefont {Y.}~\bibnamefont {Nisikawa}},\ and\
  \bibinfo {author} {\bibfnamefont {K.}~\bibnamefont {Yamada}},\ }\bibfield
  {title} {\bibinfo {title} {Theory of superconductivity in {CeMIn$_5$ (M=Co,
  Rh, Ir)} on the basis of the three dimensional periodic anderson model},\
  }\href {https://doi.org/10.1143/JPSJ.75.024713} {\bibfield  {journal}
  {\bibinfo  {journal} {Journal of the Physical Society of Japan}\ }\textbf
  {\bibinfo {volume} {75}},\ \bibinfo {pages} {024713} (\bibinfo {year}
  {2006})}\BibitemShut {NoStop}%
\bibitem [{\citenamefont {Coleman}(2015)}]{Coleman_2015}%
  \BibitemOpen
  \bibfield  {author} {\bibinfo {author} {\bibfnamefont {P.}~\bibnamefont
  {Coleman}},\ }\href {https://doi.org/10.1017/CBO9781139020916} {\emph
  {\bibinfo {title} {Introduction to Many-Body Physics}}}\ (\bibinfo
  {publisher} {Cambridge University Press},\ \bibinfo {year}
  {2015})\BibitemShut {NoStop}%
\bibitem [{\citenamefont {Kontani}\ \emph {et~al.}(2009)\citenamefont
  {Kontani}, \citenamefont {Goryo},\ and\ \citenamefont
  {Hirashima}}]{Kontani_PRL_2009}%
  \BibitemOpen
  \bibfield  {author} {\bibinfo {author} {\bibfnamefont {H.}~\bibnamefont
  {Kontani}}, \bibinfo {author} {\bibfnamefont {J.}~\bibnamefont {Goryo}},\
  and\ \bibinfo {author} {\bibfnamefont {D.~S.}\ \bibnamefont {Hirashima}},\
  }\bibfield  {title} {\bibinfo {title} {Intrinsic spin hall effect in the
  $s$-wave superconducting state: Analysis of the rashba model},\ }\href
  {https://doi.org/10.1103/PhysRevLett.102.086602} {\bibfield  {journal}
  {\bibinfo  {journal} {Phys. Rev. Lett.}\ }\textbf {\bibinfo {volume} {102}},\
  \bibinfo {pages} {086602} (\bibinfo {year} {2009})}\BibitemShut {NoStop}%
\bibitem [{\citenamefont {Biderang}\ and\ \citenamefont
  {Yavari}(2016)}]{Biderang_PhyisicaC_2016}%
  \BibitemOpen
  \bibfield  {author} {\bibinfo {author} {\bibfnamefont {M.}~\bibnamefont
  {Biderang}}\ and\ \bibinfo {author} {\bibfnamefont {H.}~\bibnamefont
  {Yavari}},\ }\bibfield  {title} {\bibinfo {title} {Spin hall conductivity in
  the impure two-dimensional rashba s-wave superconductor},\ }\href
  {https://doi.org/10.1016/j.physc.2016.03.011} {\bibfield  {journal} {\bibinfo
   {journal} {Physica C: Superconductivity and its Applications}\ }\textbf
  {\bibinfo {volume} {525-526}},\ \bibinfo {pages} {100–104} (\bibinfo {year}
  {2016})}\BibitemShut {NoStop}%
\bibitem [{\citenamefont {Bonalde}\ \emph {et~al.}(2005)\citenamefont
  {Bonalde}, \citenamefont {Brämer-Escamilla},\ and\ \citenamefont
  {Bauer}}]{Bauer_PRL_2005}%
  \BibitemOpen
  \bibfield  {author} {\bibinfo {author} {\bibfnamefont {I.}~\bibnamefont
  {Bonalde}}, \bibinfo {author} {\bibfnamefont {W.}~\bibnamefont
  {Brämer-Escamilla}},\ and\ \bibinfo {author} {\bibfnamefont
  {E.}~\bibnamefont {Bauer}},\ }\bibfield  {title} {\bibinfo {title} {Evidence
  for line nodes in the superconducting energy gap of noncentrosymmetric
  {${\mathrm{CePt}}_{3}\mathrm{Si}$} from magnetic penetration depth
  measurements},\ }\href {https://doi.org/10.1103/PhysRevLett.94.207002}
  {\bibfield  {journal} {\bibinfo  {journal} {Phys. Rev. Lett.}\ }\textbf
  {\bibinfo {volume} {94}},\ \bibinfo {pages} {207002} (\bibinfo {year}
  {2005})}\BibitemShut {NoStop}%
\end{thebibliography}%
	
\end{document}